\documentclass[12pt,a4paper,dvips]{iopart}
\usepackage{graphicx}
\usepackage{iopams}

\begin{document}
\begin{flushright} \today \end{flushright}
\jl{2} \title[Resonant 3--photon ionization \ldots] {Resonant
  3--photon ionization of hydrogenic atoms by non-monochromatic laser
  field}

\author{V Yakhontov\dag \footnote[2]{On leave from St.Petersburg State
    Technical University, Polytechnicheskaya 29, 195251, St.
    Petersburg, Russia. E-mail address: yakhon@maths.ox.ac.uk}, R
  Santra\S\ and K Jungmann\S}

\address{\dag\ Mathematical Institute, University of Oxford, St.
  Giles' 24-29, Oxford OX1 3LB, UK, and at} \address{\S\ 
  Physikalisches Institut der Universit\"{a}t Heidelberg,
  Philosophenweg 12, D-69120 Heidelberg, Germany}

\begin{abstract}
  We present ionization probability and line shape calculations for
  the two-step 3-photon ionization process, $1S \stackrel{2\hbar
  \omega}{\longrightarrow}2S \stackrel{\hbar
  \omega}{\longrightarrow}\varepsilon P $, of the ground state of
  hydrogenic atoms in a non-monochromatic laser field with a
  time--dependent amplitude. Within the framework of a three--level
  model, the {\it AC Stark \/} shifts and non-zero ionization rates of
  all states involved were taken into account together with spatial
  and temporal inhomogeneities of the laser signal. In contrast with
  the usual perturbative technique, the time evolution of the atomic
  states was simulated    by direct numerically solving the system of
  coupled time--dependent inhomogeneous differential equations, being
  equivalent to the appropriate non-stationary Schr\"{o}dinger
  equation.  Particular numerical results were obtained for typical
  parameters of the pulsed laser field that are employed in a new
  experiment to measure the $1S-2S$ energy separation in muonium at
  the Rutherford Appleton Laboratory. The shifts and asymmetries of
  the photoionization line shapes revealed may be of relevance for
  ultra-high precision experiments in hydrogen in CW laser fields.  
  \end{abstract}
 \submitted \maketitle

\section{Introduction}
\label{sec:int}

The dynamics of relaxing quantum systems in sufficiently strong laser
fields, which are neither stationary nor monochromatic, is an
important aspect of the theory of interaction between atoms and photon
fields.  This problem is of particular practical significance because
of dramatic advances in the precision of measurements presently
attainable in spectroscopic experimental studies of hydrogenic and
few--particle atoms.  Among the simple atoms that attracted attention
in the last few years one should mention: {\em hydrogen} and its
isotopes \cite{Nez,Bourzeix,Berkeland,Schmidt,Udem}, {\em positronium}
\cite{Fee,Mills}, denoted $(e^+-e^-)$, {\em muonium} (see
\cite{Maas,Jung,Mills,Putlitz,Boshier} and references therein),
denoted $(\mu^+-e^-)$, and {\em helium} atom \cite{Eikema}.  The
measurements of the ground state HFS and the Lamb shift are of
particular interest for these systems.  High precision spectroscopy of
low--lying transitions in hydrogenic atoms offers a unique opportunity
to test QED calculations and to refine both the fundamental constants
and various properties of respective nuclei.  This urges further
theoretical developments intended to describe the well established
photon-induced processes in simple atoms with much higher precision
than techniques used so far are able to provide, in order to allow a
proper interpretation of newly available experimental data.  In the
present paper the stepwise 3-photon ionization of muonium in a {\em
non--monochromatic} laser field with a {\em time--dependent} amplitude
is studied.

\begin{figure}[t]
  \centering \includegraphics[width=10cm,height=5cm]{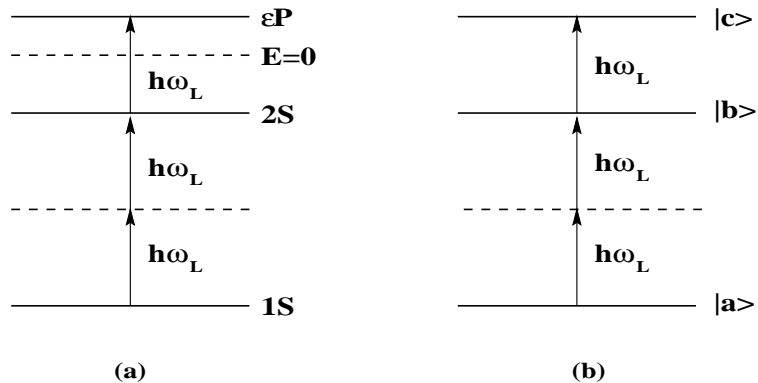}
\caption{(a) -- The scheme of the new 1S-2S experiment in
  muonium. (b) -- The set of states to model the stepwise 3--photon
  ionization.}
\label{fig1}
\end{figure}

It has been recognized already a long time ago (see \cite{Giacobino}
for a more detailed discussion) that the $1S-2S$ transition offers
unique opportunities for high precision spectroscopy due to the narrow
natural line width, $\Gamma^{\rm (nat)}_{2s}$, of the $2S$-state.  In
the hydrogen atom, for example, $\Gamma^{\rm (nat)}_{2s} \approx
1.3$~Hz which enables the quality factor of $\delta \nu/\nu \simeq
10^{-15}$ to be achieved already today, as well as suggests future
resolutions of order of 1 part in $10^{18}$.  Experimentally, the
$1S-2S$ transition can be induced Doppler--free by absorbing two
photons from two counter--propagating laser beams.  In particular,
this scheme is presently used at the Rutherford Appleton
Laboratory~\cite{Maas} for determination of the $1S-2S$ energy
separation in muonium to the $1$~MHz accuracy.  This study offers an
opportunity to improve the present knowledge of the muon/electron mass
ratio and, thereby, of the muon mass itself.  In the course of this
investigation, two counter--propagating pulsed laser beams with almost
identical intensities, $I(x,y,t),\;I_{\rm max}\approx
10^6\;\mbox{W/cm}^2$, and the same photon energies, $h \nu_{\rm
L}\approx 3/16 \;a.u.$ ($\lambda_{\rm L} = 244$ nm) are used.  Thus,
the frequency $\nu_{\rm L}$ is such that it is tuned into resonance
with the 2--photon Doppler--free $1S \stackrel{2\hbar \omega_{\rm
L}}{\longrightarrow}2S$ transition.  Unlike the hydrogen case however,
the $1S-2S$ transition in muonium is hard to be observed directly,
i.e.  by detecting radiation emitted as a result of the $2S$--state
deexcitation.  This is due to the fact that the appropriate line
intensities in muonium happen to be weaker by several orders of
magnitude, owing to much lower densities at which muonium atoms can be
produced.  Furthermore, the latter reason necessitates the use of
higher laser intensities (by a factor of $10^4$, at least) that are
required for a reasonable signal strength.  This demands a {\it
pulsed\/} rather than a CW laser source (as in the case of hydrogen
\cite{Udem}) to be employed.  The $2S$--state of muonium is detected
therefore via its photoionization by the third photon absorbed from
one of the laser beams \cite{Maas}.  Schematically, the experiment in
muonium is presented in figure~\ref{fig1}(a) where $\hbar \omega_{\rm
L} \approx 3/16\;a.u.\; (\nu_{\rm L} \equiv \omega_{\rm L}/2 \pi
\simeq 10^9\:\mbox{MHz})$ stands for the resonant energy (frequency)
which drives the 2--photon $1S-2S$ transition.

Despite the fact that the laser frequency can be calibrated, in
principle, to a rather high accuracy, there are a number of systematic
error sources each of which proves to be essential for precise
determination of the $1S-2S$ energy separation in muonium.  In this
paper we report the results of the theoretical study intended to allow
for one of the most important among these effects, that is, the
time-dependent frequency variation (chirp) of the laser field.  This
phenomenon arises due to the rapid refraction index variation of the
{\em laser media} and is hardly avoidable, especially with powerful
pulsed lasers, unless some technical developments are made in order to
compensate the frequency alteration~\cite{Maas,Eikema}.  As will be
demonstrated below, the chirped laser signal immediately makes its
appearance in appreciable spurious shifting and broadening of
spectral/photoionization lines, as well as leads to a noticeable
distortion of the line shapes.  To estimate the scale of the effect
considered, one should note, for example, that the chirp--induced
shift of the center of the ionization line is roughly equal to a
characteristic magnitude of the chirp itself. For the $1S-2S$
experiment in muonium, the latter can typically amount to $\delta/2\pi
\simeq 10 \ldots 100\:\mbox{MHz}$, thus leading to the relative shift
of the line's center by $\delta/\omega_{\rm L} \simeq 10^{-8} \ldots
10^{-7}$. The effects of such an order are usually completely
neglected in atomic physics. Nonetheless, these happen to be crucial
for studies where an absolute accuracy of $1\:\mbox{MHz}$ is
anticipated.  It is therefore of primary practical importance to work
out a relatively simple theoretical scheme which could allow one to
describe certain types of multi-photon ionization phenomena in
hydrogenic atoms, where an arbitrary time variation of the frequency
of the laser field is accompanied by an arbitrary time modulation of
its amplitude.

Although the problem of the hydrogenic systems' interaction with
monochromatic resonant laser field with/without the amplitude
modulation is investigated quite fully by now (see
\cite{Zon1,Zon2,ZMR,Giacobino,Beausoleil,Schmidt} and references
therein), there have been much fewer studies where both the amplitude
and the frequency of the laser signal vary with time.  Even though,
the authors confine themselves in most cases to either 1--photon
(rather than 2--photon) resonant transition \cite{Kaplan1}, or treat
only somewhat special forms (``abrupt step'', ``exponential field
pulses'' etc.) of the field modulation \cite{Kaplan2}.  Under these
circumstances, it seems very desirable to reconsider the problem as a
whole, while adapting it specifically to the conditions of the $1S-2S$
experiment in muonium.

\section{Theory}
\label{sec:theory}

\subsection{Basic equations}
\label{subsec:gen_cons}

We consider, without a loss of generality, a set of three levels:
$|a\rangle,\; |b\rangle$ and $|c\rangle$, of a hydrogenic atom with
the charge $Z$ of its nucleus, such that their one--particle energies
satisfy the inequality: $\varepsilon_a< \varepsilon_b <
\varepsilon_c$; this rather general setup is shown in figure
\ref{fig1}(b).  It is supposed that the given 3--level system is
exposed to two counter--propagating (along the $z$--axis) laser waves
with equal time--dependent (circular) frequencies, $\omega(t)\equiv
\omega_{\rm L}+\frac{1}{t}\phi(t)$, polarizations vectors,
$\bepsilon$, and wave vectors, $\bi{k}_1(t)=-\bi{k}_2(t)\equiv
\bi{k}(t),\;|\bi{k}(t)|= \alpha \omega(t)$, where $\alpha =e^2/(\hbar
c) \approx 1/137$ is the fine structure constant and $c$ denotes the
speed of light.  The stationary part, $\omega_{\rm L}$, of the field
frequency $\omega(t)$ is assumed to be such that it is in the 2-- and
1--photon resonances with the pairs of states,
$(|a\rangle,\:|b\rangle)$ and $(|b\rangle,\:|c\rangle)$, respectively.
This implies that $2\pi \nu_{\rm L}\equiv \omega_{\rm
  L}=\omega_{b,a}/2=\omega_{c,b}$ with $\omega_{j,i}\equiv
\varepsilon_j-\varepsilon_i,\;(i,j=a,b,c)$ being the differences of
the one--particle energies.  In addition, it is supposed that the
$E1$--transition is forbidden between $|a\rangle$ and $|b\rangle$ and
is allowed between $|b\rangle$ and $|c\rangle$.

Under the resonance conditions assumed above, 3--photon ionization of
the system, that is, $|a\rangle \stackrel{3\hbar \omega_{\rm L}}{\longrightarrow} |c\rangle$ 
transition, occurs predominantly as a
2--step process: (i) a 2--photon resonant absorption from the state
$|a\rangle$ into the state $|b\rangle$, followed by (ii) a 1--photon
resonant transition between the levels $|b\rangle$ and $|c\rangle$.
This simple physical picture is a consequence of the evident estimate
of the probability for an atom to absorb 3 photons of equal energies
(the atomic units, $e=\hbar=m_{\rm e}=1$, are used throughout the
paper; in this system, the speed of light is equal to $c \approx
137$):
\begin{equation}
\label{3phot}
W^{(3)}_{c,a}({\rm res}) \propto  \frac{\Gamma_a} {\overline{\Delta \omega}^2} 
W^{(2)}_{b,a}({\rm res}) W^{(1)}_{c,b}({\rm res}).
\end{equation}
Here, $W^{(2)}_{b,a}(\mbox{res}),\:W^{(1)}_{c,b}(\mbox{res})$ are the
2-- and 1--photon resonant transition probabilities, $\Gamma_a$ is the
total width of the state $|b\rangle$, and $\overline{\Delta \omega}$
denotes some characteristic mean energy which a two--fold summation
over intermediate states in the exact expression to define
$W^{(3)}_{c,a}(\mbox{res})$ (see \cite{LP} for a more detailed
discussion) can be reduced to. Equation (\ref{3phot}) readily follows
then from the estimate, $\Gamma_a/\overline{\Delta \omega}^2 \ll 1$,
which is fulfilled for most states of atoms in moderately strong laser
fields, except extraordinarily short--living ones.  A situation when
$|a\rangle$ stands for the ground--, $|b\rangle$ for the $2S$--, and
$|c\rangle$ for the continuum $\varepsilon P$--states, respectively,
is of our primary concern here.  In this case $\overline{\Delta
  \omega}^2\simeq 1$ and $\Gamma^{\rm (tot)}_{2s}/\overline{\Delta
  \omega}^2 \simeq 10^{-9}$ since $\Gamma^{\rm (tot)}_{2s}
=\Gamma^{\rm (nat)}_{2s} + \Gamma^{\rm (phot)}_{2s} \simeq 10^{-9}$ is
dominated by the photoionization rate of the $2S$--level ($\Gamma^{\rm
  (phot)}_{2s} \simeq 10^{-9} \gg \Gamma^{\rm (nat)}_{2s} \simeq
10^{-15}$).  Accordingly, we will assume that the natural line widths
of the $|a\rangle$- and $|b\rangle$--levels are both equal to zero:
$\Gamma^{\rm (nat)}_a= \Gamma^{\rm (nat)}_b=0$.  A straightforward
generalization of the latter condition, that might be required for the
treatment of {\it excited\/} states, can be achieved by adding
appropriate imaginary parts to the energies of the levels~\cite{LL}:
$\varepsilon_a \rightarrow \varepsilon_a - \rmi \Gamma^{\rm
  (nat)}_a/2,\;\varepsilon_b \rightarrow \varepsilon_b - \rmi
\Gamma^{\rm (nat)}_b/2$.

In the semiclassical approximation, which happens to be accurate
enough for our purposes, the electric fields inducing the 2-- and
1--photon transitions can be taken in the form
\begin{equation}
\bi{E}_{1,2}(\bi{r},t)=\frac{1}{2}\bepsilon E_{1,2}(t) U_{1,2}
(\bi{r}) \exp\left\{\rmi \left(\bi{k}_{1,2}(t) \bi{r} - \omega(t) t
\right) \right\} + c.c.
\label{fields}
\end{equation} 
Here, the real functions $U_1(\bi{r})=U_2(\bi{r}) \equiv U(\bi{r})$
and $E_1(t)=E_2(t)\equiv E(t)$ describe the (macroscopic) spatial
inhomogeneity of the laser field and its time-dependent amplitude.
These are related direct to the laser intensity $I(\bi{r},t)$ which,
along with the chirped (circular) laser frequency $\omega(t)$, is
obtainable direct from measurements.  For the $1S-2S$ experiment with
muonium, $I(\bi{r},t)$ and $\omega(t)$ can be well approximated as
\cite{Oxford}:
\begin{eqnarray}
\label{intens}
I(\bi{r},t) & \equiv & \frac{c}{8\pi}
E^2(t)U^2(\bi{r})=\frac{1}{(2\pi)^{3/2}}\frac{A_\omega}{\sigma_{\rm t}
\sigma^2_{\rm r}}\exp
\left\{-\frac{t^2}{2\sigma_{\rm t}^2}-
\frac{x^2+y^2}{2\sigma_{\rm r}^2}\right\},\\
\label{omega}
\omega(t) & \equiv & \omega_{\rm L}+\frac{1}{t} \phi(t) 
\approx \omega_{\rm L} + \dot{\phi}(t) 
\equiv \omega_{\rm L} + \frac{1}{2} \delta 
\left( 1+ {\rm erf}(t/\tau)\right),
\end{eqnarray}
where $ A_{\omega }=2\ldots 5$~mJ is the energy within one laser
pulse; $\sigma_{\rm t}=40 \ldots 65$~ns, $\sigma _{\rm r}=0.5\ldots
1.5 $~mm denote the temporal and spatial dispersions of the external
laser field (in the $ XOY$ plane perpendicular to the direction of the
beams' propagation).  The full temporal width of the laser pulse is
expressed then as $\tau=2\sqrt{2\ln 2}\cdot \: \sigma_{\rm t}
=100\ldots 150$~ns. Parameter $\delta \equiv \omega_{\rm L}(+\infty) -
\omega_{\rm L}(-\infty) \simeq 2\pi \cdot (10 \ldots 100)$ MHz defines
the magnitude of the chirp $\dot{\phi}(t)$, being actually detected in
the experiment, and $\mbox{erf}(\ldots) $ stands for the error
function \cite{AS}.

As usually happens to be true in most practical situations, both the
amplitude $E(t)$ and the additonal phase $\phi(t)$ vary much slower
with time than $\bi{E}(t)$ and $\omega(t)$(see equations
(\ref{intens})-(\ref{omega})).  In addition, the magnitude of the
chirp is normally much smaller than the ``base'' frequency
$\omega_{\rm L}$. It is therefore permissible to assume that
$\omega(t)$ and $E(t)$ are subject to the following general conditions
\cite{CT1968,Levinson}:
\begin{eqnarray}
\label{cond_omega}
|t \dot{\omega}(t)| \ll |\omega(t)-\omega_{\rm L}|
\equiv \left| \frac{1}{t}\phi(t) \right| \approx |\dot{\phi}(t)| \ll
\omega_{\rm L} \\
\label{cond_E}
\left| \dot{E}(t) \right| \ll \left| \omega_{\rm L}
E(t) \right|.
\end{eqnarray}
Within the framework of the conventional time--dependent perturbation
theory \cite{LL}, the exact time--dependent wave function of the
quasi--stationary state $\psi_n(t)$ in the field is sought in the form
\begin{equation}
\psi_n(t)=\sum_{k} c_{k,n}(t) \psi^{(0)}_{k} \e^{-\rmi \varepsilon_k t},
\label{exp_psi}
\end{equation} 
where $c_{k,n}(t)$, being the amplitudes to find an atom in one of the
unperturbed states $\psi^{(0)}_{k}$, satisfy the following well known
system of coupled differential equations:
\begin{equation}
\dot{c}_{k,n}(t) =- \rmi  \sum_{s}V_{k,s}(t)c_{s,n}(t)  
\e^{-\rmi \omega_{k,s} t}.
\label{a}
\end{equation} 
Here, $V_{k,s}(t)$ denotes the matrix element of $V(t)$ being the
operator of the ``particle--laser field'' interaction. Due to the
given below reasons, $V(t)$ can be taken in the form
\begin{equation}
V(t) \approx -\bi{E}(\bi{r},t)\bdot \bi{d} \approx
-\frac{1}{2}E(t) U(\bi{r}) \e^{-\rmi \omega(t) t} \bepsilon \bdot \bi{d} + c.c.
\label{dip_appr}
\end{equation}
where $\bi{d}$ is the dipole operator of an electron (or a muon) and
$\bi{E}(\bi{r},t)$ denotes the electric field~(\ref{fields}) seen by
the atom in its rest frame.  For particular laser parameters of
interest, the accuracy of the above approximation follows from the
evident estimate: $|\nabla U(\bi{r}) | \simeq 1/\sigma_{\rm r} \approx
10^{-7} \ll|\bi{k}(t) \bdot \bi{r}|\simeq 2\pi/\lambda_{\rm L} \approx
10^{-3}$.  This implies that the contribution of the quadrupole terms
$\propto E(t) (\nabla U(\bi{r}) \bdot \bi{d})$, which originate from
the spatial {\em macroscopic} inhomogeneity of $\bi{E}(\bi{r},t)$, is
expected to be a factor $10^{-7}$ smaller than the contribution of the
(dipole) terms in (\ref{dip_appr}).  Almost the same estimate holds
also for corrections arising due to the quadrupole component $\propto
E(t) U(\bi{r}) |\bi{k}(t)| ( \bepsilon \bdot \bi{d})^2$ of the photon
field itself. This is despite the fact that the states $|a\rangle$ and
$|b\rangle$ are coupled, in accord with the initial assumptions of our
model and equation (\ref{dip_appr}), in only the second order of the
perturbation theory in $V(t)$, i.e.  by taking into account at least
the 2--photon absorption/emission.  Indeed, unlike the second order
dipole coupling, the 1--photon quadrupole $|a\rangle \leftrightarrow
|b\rangle $ transition would apparently be off resonance. This would
makes its appearance in the extra exponential factor $\simeq \e^{\rmi
  \omega_{\rm L} t}$ in the amplitude.  If $V_{a,b} \ll \omega_{\rm
  L}$, as it is assumed throughout the paper, then the terms of such a
type are known to be almost negligible \cite{Salzman}. In our case, in
particular, their relative contribution is of the order of
$1/(\omega_{\rm L} \tau) \simeq 10^{-8}$, as can be shown by averaging
the transition amplitude over some time interval: $- T < t < T\simeq
\tau$.  The same ``off-resonant'' argument is fully applicable to
magnetic transitions as well, thus elucidating why the $M1$-- and
$E2$--contributions are suppressed by a factor $\simeq 10^{-8}$ as
compared to the dipole one.

By virtue of (\ref{a})-(\ref{dip_appr}), one can readily retrieve,
after some conventional algebra, the following form for effective
quasiclassical one--particle operators which couple two given states
of reference \cite{Beausoleil,Giacobino}:
\begin{eqnarray}
\label{Vba}
V^{\rm (eff)}_{b \leftarrow a}
(\bi{r},t,\omega_{\rm L}) & = &
\frac{1}{4}U^2(\bi{r}) E^2(t) \e^{-2 \rmi (\omega_{\rm L} t +\phi(t))}
 \sum_{s} \hspace{-14.5pt} \int 
\frac{\langle b|\bi{d} \bdot \bepsilon|s \rangle 
\langle s|\bi{d} \bdot \bepsilon|a \rangle } 
{\omega_{a,s} +\omega_{\rm L} }, \\
\label{Vab}
V^{\rm (eff)}_{b \rightarrow a}
(\bi{r},t,\omega_{\rm L}) & = &
\frac{1}{4}U^2(\bi{r}) E^2(t) \e^{2 \rmi (\omega_{\rm L} t +\phi(t))}
 \sum_{s} \hspace{-14.5pt} \int 
\frac{\langle a|\bi{d} \bdot \bepsilon^\ast |s \rangle 
\langle s|\bi{d} \bdot \bepsilon^\ast |b \rangle } 
{\omega_{b,s} -\omega_{\rm L} } , \\
\label{Vbb}
V^{\rm (eff)}_{b,b}(\bi{r},t,\omega_{\rm L}) & = &
\frac{1}{4}U^2(\bi{r}) E^2(t) 
 \sum_{s} \hspace{-14.5pt} \int \left \{ 
\frac{|\langle b|\bi{d} \bdot \bepsilon |s \rangle|^2} 
{\omega_{b,s} +\omega_{\rm L}} + 
\frac{ |\langle b|\bi{d} \bdot \bepsilon |s \rangle |^2}
  {\omega_{b,s} - \omega_{\rm L} } \right\}. 
\end{eqnarray}
All summations are performed here over complete set of one-particle
atomic orbitals.  Equations (\ref{Vba})-(\ref{Vab}) describe the real
transitions, whereas (\ref{Vbb}) defines the {\it AC Stark\/} shift of
the level $|b\rangle$ (or $|a\rangle$, if $b \rightarrow a$) which
arises due to interaction of the atom with the incident radiation.
Note that the sum in (\ref{Vbb}) coincides, by definition \cite{LL},
with the dipole dynamic polarizability of the state $|b\rangle$ at the
frequency $\omega$.  Also, it should be pointed out that only the
leading terms, as given by equations (\ref{Vba})-(\ref{Vbb}), were
retained in the course of derivation.  The most important among
neglected terms are of the form:
\[
\fl U^2(\bi{r}) \left.  \left| \frac{\rmi}{\omega_{s,a}-\omega_{\rm
        L}} ( \frac{\dot{E}(t)- \rmi E(t) \dot{\phi}(t)}
    {\omega_{s,a}-\omega_{\rm L}} \e^{ \rmi \left[
        (\omega_{s,a}-\omega_{\rm L})t -\phi(t)\right] } + c.c.
  \right ) \right|^2 \simeq \frac{1}{\omega_{\rm L}^4}
\left|\dot{E}(t)\right|^2U^2(\bi{r}).
\]
Apparently, this contribution is smaller than $V^{\rm (eff)}_{i
  \leftrightarrow j}(\bi{r},t,\omega_{\rm L}),\;(i,j=a,b)$ by a factor
of $\left|\dot{E}(t)/(\omega_{\rm L} E(t))\right|^2 \simeq
1/(\omega_{\rm L} \tau) \simeq 10^{-8}$, as follows from
(\ref{intens})-(\ref{omega}), in accord with
(\ref{cond_omega})-(\ref{cond_E}).

As was assumed above, there are no other discrete resonant atomic
levels (either 1-- or 2--photon) except $|a\rangle,\; |b\rangle$.
This enables the system (\ref{a}) to be reduced to these two states
only \footnote{This implies that the Coulomb degeneracy of the levels
  with the same principal quantum number, say $|ns\rangle$ and
  $|nd\rangle$, is ignored, since we are mainly concerned here with
  the lowest $1S$- and $2S$--states. }, while describing the 1st stage
of the two--step 3--photon ionization process under consideration.
Indeed, an inclusion of the off-resonant states would give rise in the
right-hand side of equations (\ref{ala1})-(\ref{CC1}) to the rapidly
$t$--varying terms $\simeq \e^{\pm i\omega_{\rm L} t}$. After temporal
averaging (see above), these happen to be of order of $1/(\omega_{\rm
  L} \tau) \simeq 10^{-8}$ , and can therefore be omitted to the
accuracy adopted in this study.  On introducing the new functions,
\begin{equation}
\fl c_{a,a}(t)\equiv \e^{- \rmi \alpha(t)},
\qquad C(t)\equiv c_{b,a}(t)
\exp\left\{ \rmi \left[2\left (\omega-\frac{1}{2}
\omega_{b,a}\right) +
\alpha(t) +2 \phi(t)\right] \right\},
\label{alpha_C}
\end{equation}
and replacing the matrix elements in (\ref{a}) by equations
(\ref{Vba})-(\ref{Vbb}), one arrives at the following system of
equations:
\begin{eqnarray}
\label{ala1}
\fl \dot{\alpha}(t) & = & \frac{4\pi}{c} I(\bi{r},t)  \left[
D_{a,a} +  D_{a,b} C(t) \right], \\ \fl
\label{CC1}
\dot{C}(t) & = & - \rmi \:D_{b,a}  \frac{4\pi}{c} 
I(\bi{r},t) - \rmi \: \left [ D_{b,b}
\frac{4\pi}{c} I(\bi{r},t) -
\left( 4\pi \Delta \nu + \dot{\alpha}(t)+2
\dot{\phi}(t)\right)
\right] C(t),
\end{eqnarray}
where $2\pi \Delta \nu \equiv \omega_{\rm L}-\omega_{b,a}/2$ is the
time--independent frequency detuning off the resonance, $I(\bi{r},t)=
c E^2(t)U^2(\bi{r})/8\pi$ is the intensity of the single laser beam
and $D_{j,i},\;(j,i)=a,b$ denote the effective matrix elements given
by the sums in (\ref{Vba})-(\ref{Vbb}).  It has been taken into
account here that the 3--level system is actually probed by a
superposition of two laser fields (\ref{fields}), rather than by a
single one; this results eventually in doubling all matrix elements.
In addition, we have neglected a contribution due to the 2--photon
absorption from {\em each} of two laser beams.  This contribution
comes from the second--order interactions of an atom with only one of
two light waves and makes its appearance finally in two
Doppler--broadened terms, i.e.  depending on a $v_z$--component of
atomic velocity in the laboratory frame, in the expression to define
ionization probabilities.  However, after averaging over $v_z$, as
would be required within the framework of a systematic approach, these
terms turn out to be small compared, at the center of the
photoionization line \cite{Chebotaev, Bloembergen}, with the
Doppler--free ones which originate from (\ref{ala1})-(\ref{CC1}).
These arguments provide the ground for the approximation made,
although the effects caused by atomic movement in the media are beyond
the scope of our present consideration.

It is relevant to point out here that the matrix elements,
$D_{j,i},\;(j,i)=a,b$, in (\ref{ala1})-(\ref{CC1}) are $\omega_{\rm
  L}$--dependent.  In particular, $D_{b,b}$ has a non--zero imaginary
part. The latter allows for a non--zero photoionization rate of the
$|b\rangle$--state, due to the action of the photon field, since
$\omega_{\rm L}$ is supposed to exceed the photoionization threshold
$I_b$.  Indeed, as was noted above, $D_{b,b}$ is proportional to the
dynamic (tensor) polarizability $\alpha_b^{ij}(\omega_{\rm L})$ of the
level $|b\rangle$, which is complex-valued if $\omega_{\rm L} \geq
I_b$ (see our recent paper \cite{YJ} for a more detailed discussion).
$\Re\alpha_a^{ij}(\omega)$ and $\Im\alpha_a^{ij}(\omega)$ define then
the {\it AC Stark\/} shift of the level (see equation (\ref{Vbb})) and
its decay probability ($\propto |V_{bc}|^2$) via the single
photoionization. This result is in agreement with the optical
theorem~\cite{LP}: $\sigma_b^{(\gamma)}(\omega_{\rm L})=
4\pi\alpha\omega_{\rm L} \Im\alpha_b^{ij}(\omega_{\rm L})$, where
$\alpha$ is the fine structure constant and
$\sigma_b^{(\gamma)}(\omega_{\rm L})$ denotes the total
photoionization cross section of the state $|b\rangle$.  Numerical
calculations show that the single photoionization happens to be the
main channel of the $|b\rangle$--level depopulation. It is therefore
crucial for the correct description of the time--evolution of the
3--level system that this depopulation mechanism is taken in fact into
account in (\ref{ala1})-(\ref{CC1}) through the matrix elements
$D_{j,i},\;(j,i)=a,b$.

An allowance for the 2nd step of the process considered can be made by
augmenting the system (\ref{ala1})-(\ref{CC1}) with an additional
equation that explicitly describes the $|b\rangle - |c\rangle$
coupling through the single photoionization of the state $|b\rangle$.
Given by (\ref{cont}), this extra equation is nothing else but a
differential form of the conservation law which determines the
probability balance between the states of reference. Moreover,
equations (\ref{ala1})-(\ref{CC1}) can be decoupled by substituting
$\dot{\alpha}(t)$ of (\ref{ala1}) into (\ref{CC1}) to finally get
\numparts
\begin{eqnarray}
\label{CC2}
\fl \dot{C}(t) =  - \rmi D_{b,a} \frac{4\pi}{c} 
I(\bi{r},t) + 
  \rmi \left [\left( D_{a,a}- D_{b,b}\right) \frac{4\pi}{c} I(\bi{r},t) + 
2 \left( 2 \pi \Delta \nu + \dot{\phi}(t)\right) \right] C(t)  \\ 
   \lo+  \rmi D_{b,a} \frac{4\pi}{c} 
I(\bi{r},t) C^2(t), \nonumber  \\
\label{ala2}
\fl \dot{\alpha}(t) =   \frac{4\pi}{c} I(\bi{r},t) \left [
D_{a,a} + D_{b,a} C(t) \right],\\
\label{cont}
 \fl \dot{W}_c(t) = 
\frac{2}{\omega_{\rm L}}
 \sigma^{(\gamma)}_b
(\omega_{\rm L})
I(\bi{r},t) \left| C(t)\right|^2 
\exp \left\{2\Im  \alpha(t) \right\},
\end{eqnarray}
\endnumparts where $W_c(t)$ is the probability for an atom to be
ionized by the time $t$, via absorption of 3 photons of equal energy.
Under the actual experimental conditions \cite{Maas}, initial
population of the level $|b\rangle$, i.e.  before the laser pulse is
shot, is usually rather low. This is because the major fraction of the
muonium atoms is produced in the {\em ground} state $|a\rangle$ where
the population is proportional to $|c_{a,a}|^2\equiv \exp \left\{2\Im
  \alpha(t) \right\}$.  In view of this physical picture and by virtue
of (\ref{alpha_C}), it is natural to impose the following initial
conditions on the functions $\alpha(t),\:C(t),\: W_c(t)$:
\begin{equation}
C(-\infty)=0, \qquad \alpha(-\infty)=0, \qquad W_c(-\infty)=0.
\label{cond_inf} 
\end{equation} 
These enable $W_c(t=+\infty,\Delta \nu,A_\omega,r;[\phi])$ to be
determined uniquely. Being of our primary concern, this quantity
constitutes the required 3--photon resonant ionization probability of
an atom at infinitely large positive times, i.e. when the interaction
between the single laser pulse and the system has already ceased. As a
function of the laser frequency detuning $\Delta \nu$,
$W_c(t=+\infty,\Delta \nu,A_\omega,r;[\phi])$ describes, and will be
therefore used as a synonym of, the 3--photon ionization line
profile/shape. For brevity, the latter will be denoted as
$W_c^{\infty}(\Delta \nu,\delta,A_\omega,r;[\phi])$ where the argument
$[\phi]$ is introduced in order to indicate explicitly that the line
profile depends on the form of the chirp. In particular, if
$\dot{\phi}(t) \equiv \delta \left(1+ {\rm erf}(t/\tau)\right)/2$ then
$W_c^{\infty}(\Delta \nu,A_\omega,r;[\phi])$ will be denoted as simply
$W_c^{\infty}(\Delta \nu,A_\omega,\delta,r)$.

Equations (\ref{CC2})-(\ref{cont}) are well suited for numerical
calculations with arbitrary functions defining the spatial and
temporal distributions of the laser pulse.  In addition, the given
system is rather convenient for analytic treatment, as will be
demonstrated in subsection \ref{subsec:anal_cons}. In particular, one
can develop further perturbative expansion of equations
(\ref{CC2})-(\ref{ala2}) in terms of the intensity $I(\bi{r},t)$.
This yields analytic formulae which, after setting $\dot{\phi}(t) = 0
$, agree with those formerly derived in \cite{Beausoleil} within the
framework of a 2--level model, for {\em unchirped} laser signals.
Despite a relatively simple form of the appropriate equations,
however, a not very straightforward numerical integration is still
required in order to calculate, for example, photoionization line
profiles in most practical situations, let alone the fact that the
result of the work \cite{Beausoleil} is valid for rather weak laser
intensities only.  Note that equations (\ref{CC2})-(\ref{cont}) are
free from the latter restriction.  Furthermore, unlike our approach
employing the probability amplitudes, the problems similar to those
considered in this paper are usually treated (e.g. see
\cite{Beausoleil,Eikema}) in the formalism of the density matrix
\cite{Scully}.  Under the physical conditions adopted here, these two
approaches are equivalent. To facilitate the adequate comparison
however, it is relevant to present here an alternative form of
(\ref{CC2})-(\ref{cont}), by rewriting the system in terms of the
density matrix elements: $\rho_{a,a}(t) \equiv |c_{a,a}(t)|^2,\:
\rho_{b,b}(t) \equiv |c_{b,a}(t)|^2$ and $\rho_{b,a}(t) \equiv
c_{b,a}(t) c_{a,a}^\ast(t) \exp\{- \rmi \:2 (2 \pi \Delta \nu t
+\phi(t))\} $, denoted as $\rho_{b,a}(t) \equiv \gamma(t) + \rmi
\beta(t)$.  This yields: \numparts
\begin{eqnarray}
\label{r2s}
\fl \dot{\rho}_{b,b}(t) = \frac{8 \pi}{c} 
\Im D_{b,b}  I(\bi{r},t) \rho_{b,b}(t) - 
     \frac{8 \pi }{c} D_{b,a}\: I(\bi{r},t) \beta(t) \\
\fl \dot{\beta}(t) = 2 \left(2\pi \Delta \nu + \dot{\phi}(t)\right) \gamma(t) +
     \frac{4 \pi}{c} I(\bi{r},t) \left\{
    \left(\Re D_{a,a} - 
\Re D_{b,b} \right) \gamma(t) \right. \nonumber \\
\label{r_beta} 
   \lo +  \left. \left(\Im D_{a,a} + 
\Im D_{b,b}\right) \beta(t) \right\}
       + \frac{4 \pi}{c} D_{b,a}\:I(\bi{r},t) 
    \left(\rho_{b,b}(t)- \rho_{a,a}(t)\right) \\
\label{r1s}
\fl \dot{\rho}_{a,a}(t) = \frac{8 \pi}{c} 
\Im D_{a,a}\:  I(\bi{r},t) \rho_{a,a}(t) 
     + \frac{8 \pi}{c} D_{b,a}\: I(\bi{r},t) \beta(t) \\
\fl \dot{\gamma}(t) =  -2 \left(2 \pi \Delta \nu + \dot{\phi}(t)\right) \beta(t) +
    \frac{4 \pi}{c} I(\bi{r},t) \left\{
    \left(\Re D_{b,b} - 
\Re D_{a,a} \right) \beta(t) \right. \nonumber \\               
\label{r_gama}   
    \lo+  \left. \left(\Im  D_{a,a} + 
\Im D_{b,b} \right)  \gamma(t) \right\} \\
\label{r_cont}    
\fl \dot{W}_c(t) =  \frac{2}
{\omega_{\rm L}} 
\sigma^{(\gamma)}_b(\omega_{\rm L})
                   I(\bi{r},t) \rho_{b,b}(t).     
\end{eqnarray}
\endnumparts Here, 5 unknown functions are subject to the following
initial conditions:
\begin{equation}
\fl \rho_{a,a}(-\infty)=1,
\qquad \rho_{b,b}(-\infty)=\gamma(-\infty)=
\beta(-\infty)=W_c(-\infty)=0.
\end{equation}
It should be pointed out that only real quantities enter equations
(\ref{r2s})-(\ref{r_cont}). This circumstance may be advantageous,
especially for numerical calculations, since most standard numerical
packages are not applicable direct to systems of ODE involving
complex--valued coefficients. Apart from that,
(\ref{CC2})-(\ref{cont}) and (\ref{r2s})-(\ref{r_cont}) generate
identical results, as has been proved by extensive numerical tests
carried out by us for laser signals with various forms of the chirp,
pulse envelope, and spatial distribution.
However, before we demonstrate appropriate results (see section
\ref{sec:appl}), it is instructive to present here an approximate but
reasonably accurate analytic solution for $C(x)$. This consideration
seems to be rather useful for a qualitative description of the process
as a whole, since it reveals some key features which make their
appearance due to the non-stationarity of both the frequency and the
amplitude of the laser signal.

\subsection{Some analytic consideration:  equation (\ref{CC2}) }
\label{subsec:anal_cons}

Let us assume, in agreement with (\ref{intens}), that the intensity of
the incident radiation is Gaussian both in time and space and use the
most general form of the chirp, $\dot{\phi}(t/\tau)$ (cf. equation
(\ref{omega})).

To start with, consider Riccati--type equation (\ref{CC2}) to
determine the complex--valued function $C(x):\:|C(x)| \equiv
|c_{b,a}(t)| /|c_{a,a}(t)|$, defined by (\ref{alpha_C}).  On
introducing the dimensionless variable $x \equiv t/\tau$, equation
(\ref{CC2}) takes the form
\begin{eqnarray}
\fl C'(x)- \rmi D_{b,a} {\cal I} 
\e^{-\frac{\tau^2 x^2}{2 \sigma^2_{\rm t}} }C^2(x) -
\rmi \left[ 4\pi \Delta \nu \tau +2 \dot{\phi}(x) +
{\cal I} \Delta D \e^{-\frac{\tau^2 x^2}{2 \sigma^2_{\rm t}}} \right] C(x) 
\nonumber \\
\label{eq:full}
\lo+  \rmi D_{b,a} {\cal I} \e^{-\frac{\tau^2 x^2}{2 \sigma^2_{\rm t}}}=0,
\end{eqnarray}
where the following notations were introduced for brevity:
\begin{equation}
\label{defin:Iab}
\eqalign{ 
\frac{4\pi}{c}\tau I(\bi{r},t) \equiv {\cal I} 
        \exp \left\{-\frac{\tau^2 x^2} {2 \sigma^2_{\rm t}}\right\},\qquad 
{\cal I} \equiv \sqrt{\frac{2}{\pi c^2}}\frac{\tau A_\omega}
{\sigma_{\rm t} \sigma^2_{\rm r}}
\exp \left\{-\frac{r^2}{2\sigma_{\rm r}^2}\right\}, \\
\Delta D  \equiv D_{a,a}-D_{b,b} .}
\end{equation}
Note that
\[
\fl {\cal I}= \sqrt{2\pi} \frac{2\tau}{c \sigma_{\rm t}}
\int_{-\infty}^{+\infty} I(\bi{r},t) \rmd t =
8.805\cdot 10^{-3} \frac{(A_\omega/1\:{\rm mJ})}
{(\sigma_{\rm r}/1\:{\rm mm})^2}
\exp \left\{-\frac{r^2}{2\sigma_{\rm r}^2}\right\} \:a.u.
\]
depends on the time--independent parameters of the pulse only, but not
on $x$.

\begin{figure}[t]
  \centering
  \includegraphics[height=8cm,width=8cm]{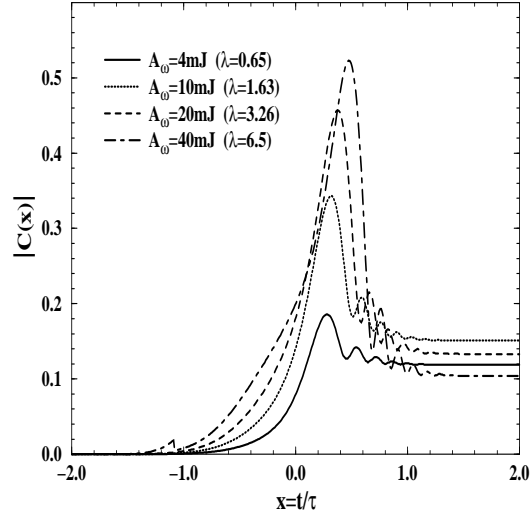}
\caption{$|C(x,\Delta \nu,\delta,A_\omega)| $ as a function of $x$,
  for various pulse energies, $A_\omega$, and fixed values of the
  chirp, $\delta =500\:{\rm Mrad}\cdot {\rm s}^{-1}$, and the
  frequency detuning, $\Delta \nu=-\delta/(4\pi)=-250/(2\pi)\:{\rm
  MHz}$. The laser field frequency is assumed to be of the form:
  $\nu(t) \equiv \nu_{\rm L}+\nu_{\rm chirp}(t) \equiv \nu_{\rm L} +
  \frac{\delta}{4\pi} \left(1+{\rm erf}(t/\tau)\right)$; the rest
  laser parameters used are: $\sigma_{\rm t}=51\:{\rm ns},\;
  \tau=120\:{\rm ns},\;r=0$.}
\label{fig1a}
\end{figure}

Due to its non-linearity, equation (\ref{eq:full}) admits only
numerically solving, except for the trivial case, $\Delta
\nu=\dot{\phi}(x)=0$, where $C(x)$ is of the form
\begin{eqnarray}
\fl C(x) = \frac{1}{2  D_{b,a}} \Biggl\{ 
\sqrt{4 D^2_{b,a}+\Delta D^2} \,
\tanh \biggl[
- \rmi \frac{\sqrt{2 \pi }\, {\cal I} \, 
        \sigma_{\rm t}}{4\tau} \sqrt{4 D^2_{b,a}+\Delta D^2}
\left(1+{\rm erf} \left(\frac{\tau x}{\sqrt{2}\sigma_{\rm t}}\right) \right)
          \nonumber \\
\lo{+} \tanh^{-1} \left( 
\frac{\Delta D}{\sqrt{4 D^2_{b,a}+\Delta D^2}} \right) \biggr] -
\Delta D \Biggr\}.
\label{C0}
\end{eqnarray}
However, the problem can be simplified considerably by linearizing
equation (\ref{eq:full}). This approximation, which the consideration
developed below is stemmed on, is justifiable if $|C(x)|^2 \ll |C(x)|
\ll 1$ holds uniformly for $-\infty < x < +\infty$ and arbitrary
$\Delta \nu,\;\delta$ and $r$. To satisfy the latter condition, it is
sufficient (but not necessary) to assume that
\begin{equation}
\label{cond_C}
\lambda \equiv \left|{\cal I} D_{b,a} (\omega_{\rm L}) \right| \lesssim 1,
\end{equation}
as readily follows from equation (\ref{C:simp}).  If $|a\rangle =
1s,\:|b\rangle = 2s$ and $r=0$, for example, then $\lambda \simeq
A_\omega |D_{2s,1s}(\omega_{\rm L})| / \sigma_{\rm r}^2 \lesssim 1$. This
imposes the following restriction on the pulse energy: $A_{\omega}
\lesssim 6\:\mbox{mJ}$, where particular values of the atomic matrix
element, $D_{2s,1s}(\omega_{\rm L})$ (see table \ref{tabD}), and the
typical laser parameters have been used.  The above estimates are
illustrated in figure \ref{fig1a} where $|C(x,\Delta
\nu=-\delta/4\pi,\delta,A_\omega)|$ obtained by direct numerically
solving equation (\ref{eq:full}) is plotted versus the dimensionless
variable $x$, at several $A_\omega$--values.  In addition, for each
pulse energy, the appropriate $\lambda$--value defined by equation
(\ref{cond_C}) is indicated for reference as well.

\begin{figure}[t]
  \centering \includegraphics[height=8cm,width=8cm]{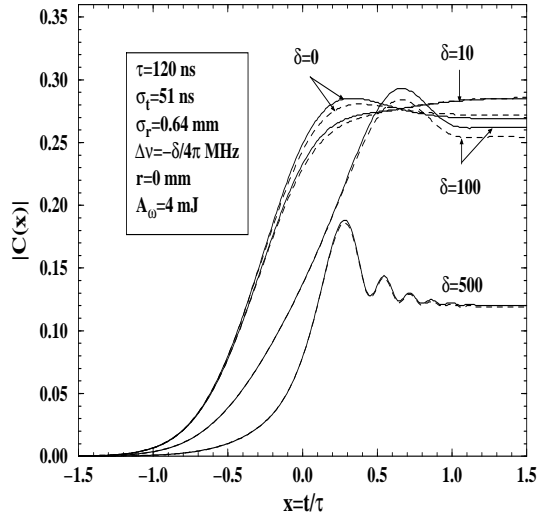}
\caption{$|C(x,\Delta \nu,\delta,A_\omega)|$ and
  $|\widetilde{C}(x,\Delta \nu,\delta,A_\omega)|$ as functions of $x$,
  for fixed values of the pulse energy, $A_\omega=4\:{\rm mJ}$,
  frequency detuning, $\Delta \nu = -\delta/(4\pi)\:{\rm MHz}$, and
  for several chirp values: $\delta =0,\:10,\:100,\:500\:{\rm
  Mrad}\cdot {\rm s}^{-1}$. The frequency of the non-monochromatic
  laser field is assumed to be of the form: $\nu(t) \equiv \nu_{\rm
  L}+\nu_{\rm chirp}(t) \equiv \nu_{\rm L} + \frac{\delta}{4\pi }
  \left(1+\mbox{erf}(t/\tau)\right)$, and the rest laser parameters
  used are: $\sigma_{\rm t}=51\:{\rm ns},\; \tau=120\:{\rm
  ns},\;r=0$.\\ $\full$ -- numerical solution of equation
  (\ref{eq:full}); $\dashed$ -- analytic solution,
  $|\widetilde{C}(x)|$, of equation (\ref{eq:simp}), as given by
  (\ref{C:simp}).}
\label{fig1b}
\end{figure}

The linearized form of equation (\ref{eq:full}) reads:
\begin{eqnarray}
\label{eq:simp}
\fl \widetilde{C}'(x)-\rmi \left[ 4\pi \Delta \nu \tau +2 \dot{\phi}(x)   + 
{\cal I} \Delta D \e^{-\frac{\tau^2 x^2}{2 \sigma^2_{\rm t}}} \right] 
\widetilde{C}(x) +
\rmi D_{b,a}\: {\cal I} \e^{-\frac{\tau^2 x^2}{2 \sigma^2_{\rm t}}}=0.
\end{eqnarray}
Its solution obeying the zero initial condition (\ref{cond_inf}) at
$x=-\infty$ has the form
\begin{eqnarray}
\label{C:simp}
\fl \widetilde{C}(x)=- \rmi D_{b,a} {\cal I} \exp \left\{
\rmi  x \left(4\pi\tau \Delta \nu + 2 \phi(x)/x  \right) +
\rmi \sqrt{\frac{\pi}{2}} \frac{\sigma_{\rm t}}{\tau}\,
{\cal I} \Delta D {\rm erf}\left(
\frac{\tau x}{\sqrt{2} \sigma_{\rm t}}\right) \right\} \nonumber \\
\fl \times \int_{-\infty}^{x}
\exp \left\{
- \rmi  u \left(4\pi \tau \Delta \nu + 2 \phi(u)/u \right) 
- \rmi \sqrt{\frac{\pi}{2}} \frac{\sigma_{\rm t} }{\tau}\,
{\cal I} \Delta D {\rm erf}
\left( \frac{\tau u}{\sqrt{2} \sigma_{\rm t}}\right) - 
\frac{\tau^2 u^2}{2 \sigma^2_{\rm t}} \right\} \rmd u. 
\end{eqnarray}
To demonstrate the fair accuracy of this relation for $A_\omega =
4\:{\rm mJ}$ and $\phi(x) = \delta \tau x \left( 1+ {\rm
    erf}(x)\right)/2$, for example, the graph of
$|\widetilde{C}(x,\Delta \nu,\delta,A_\omega)| \simeq |c_{2s,1s}(t)|
/|c_{1s,1s}(t)|$ as a function of $x$ is presented, for several chirp
values, in figure \ref{fig1b}, along with appropriate graphs for
numerical solution of the non-linear equation (\ref{eq:full}).
Together with plots in figure \ref{fig1a}, these graphs provide
instructive information about the magnitude of $|c_{2s,1s}(t)|^2
/|c_{1s,1s}(t)|^2$ at various time moments.  In addition, both figures
clearly demonstrate a rather peculiar way in which the values for the
above ratio are (strongly) distorted due the presence of the chirp.

It is relevant to note that $\widetilde{C}(x)$ approximates the
function $C(x)$ qualitatively correctly for even those $A_\omega$
which violate condition (\ref{cond_C}). An elaborate comparison with
exact numerical solutions of (\ref{eq:full}) shows that, for most
applications, the region of the validity of equation (\ref{C:simp})
can be safely extended up to $A_\omega \lesssim 20\:{\rm mJ}$, depending
on the rest laser parameters.  At higher pulse energies, the
contribution of non--linear effects becomes so essential that the
neglect of the $\simeq C^2(x)$--term is no longer permissible. This
can be seen, for example, in figure \ref{fig1a} where, for $|x|\lesssim 1$
and $A_\omega=20\:{\rm mJ}$, appropriate non--linear contribution
amounts to 20\%.

As follows from the method of our derivation, equation (\ref{C:simp})
can be readily generalized on a wide class of similar multi-photon
ionization processes that are induced by not very intense laser fields
and which involve the chirps and laser amplitudes, such that
$|\dot{\phi}(t)| \ll \omega_{\rm L}$ and $I(\bi{r},t) \rightarrow 0$
sufficiently fast as $t \rightarrow \pm \infty$.  Note that the latter
condition ensures the rapid convergence of the integral in
(\ref{C:simp}). In addition, it makes certain that the values of all
physical quantities of interest are determined actually by only a
somewhat narrow domain of $x$ within which the laser intensity is
peaked.

Equation (\ref{C:simp}) is suitable to be used further for the
analytical study of the 3--photon ionization with muonium, where
$|a\rangle = 1s,\:|b\rangle = 2s$ and $\phi(x)\equiv \delta \tau x
\left( 1+ {\rm erf}(x)\right)/2$.  In particular, $\widetilde{C}(x)$
readily enables one to obtain analytic formulae for the probabilities:
$|c_{1s,1s}(t)|^2$, $|c_{2s,1s}(t)|^2$, and $W_c^{\infty}(\Delta
\nu,A_\omega,r;[\phi])$. However, these expressions are not given
explicitly here as they turn out to be too bulky.  Instead, the
formula (\ref{C:simp}) will be used in the following subsection
\ref{subsec:wood}, while addressing an interesting complementary
problem: a calculation of the integral of the 3--photon ionization
line profile over the entire range of the frequency detunings,
$-\infty < \Delta \nu < +\infty$.

\subsection{Further analytic consideration: the integral over the frequency detunings}
\label{subsec:wood}

In what follows, it is shown that, for given $A_\omega$ and $r$, the
integral of the 3--photon ionization line profile,
$W_c^{\infty}(\Delta \nu,A_\omega,r;[\phi])$, over the entire range of
the frequency detunings $\Delta \nu$,
\begin{equation}
\label{eq:rho_gen}
\rho(A_\omega,r;[\phi])\equiv \int_{-\infty}^{+\infty}
W_c^{\infty}(\Delta \nu,A_\omega,r;[\phi]) \rmd \Delta \nu,
\end{equation}
is almost independent of particular form of the chirp, at least for
low pulse energies.  To be more precise, equation (\ref{eq:rho_gen})
defines in fact a non-linear functional of $\phi(t)$ whose
``strength'' depends, though, on the rest laser parameters.  Within
certain range of these parameters ($A_\omega$, in the first instance),
the functional relation (\ref{eq:rho_gen}) turns out to be weak, so
that $\rho(A_\omega,r;[\phi])$ plays in this case the role of a sort
of adiabatic invariant for 3--photon resonant ionization processes
occurring under the action of the fields with slow--varying
frequencies. Hence, an ``almost--conservation'' of
$\rho(A_\omega,r;[\phi])$ provides a simple and easily verifiable
approximate criterion which enables the photoionization line profiles
corresponding to different values/forms of the chirp to be compared
and normalized.

The problem discussed here was first addressed, to our knowledge, in
\cite{Woodman}, within the framework of the similar 3--level model and
for low pulse energies only. In contrast with our present approach,
however, the genuine photoionization rate of the level $|b\rangle$
(that is, $2S$) was set in \cite{Woodman} equal, respectively, to zero
and to some time-- and energy--independent constant, while describing
the dynamics of the 1st and the 2nd stages of the resonant 3--photon
ionization. Although this approximation provides acceptable results
for $A_\omega \lesssim 6\:{\rm mJ}$, say, it is of interest, both
experimental and theoretical, to reconsider the entire problem by
releasing the above simplifying assumptions and getting, thereby, a
deeper insight into the physics of the process.

\subsubsection{\underline{Low $A_\omega$}}

Let us consider first the simplest case of low $A_\omega$, such that
condition (\ref{cond_C}) is satisfied.  In addition, we will still
assume, without a loss of generality, that the intensity of the
incident radiation is Gaussian both in time and space, without
specifying, though, particular form of the chirp for a moment.  By
making use of equation (\ref{cont}), this allows
$\rho(A_\omega,r;[\phi])$ to be expressed in terms of
$\widetilde{C}(x)$ as
\begin{eqnarray}
\fl \rho(A_\omega,r;[\phi]) = \frac{c}{4\pi}
        \frac{2}{\omega_{\rm L}} \sigma^{(\gamma)}_b {\cal I}
\int_{- \infty}^{+ \infty} \rmd \Delta \nu 
\int_{- \infty}^{+ \infty} \rmd x 
\exp \left\{- \frac {\tau^2 x^2}{2 \sigma^2_{\rm t}} \right\}
\left|\widetilde{C}(x,\Delta \nu,A_\omega,r)\right|^2
\nonumber \\
\label{eq:rho-Cappr}
\lo \times \exp \left\{ 2 D_{b,a} {\cal I} \Im 
        \left[ \int_{-\infty}^{x}   
\exp \left\{- \frac {\tau^2 u^2}{2 \sigma^2_{\rm t}} \right\}
\widetilde{C}(u,\Delta \nu,A_\omega,r) \rmd u
        \right] \right\}.
\end{eqnarray}
Here, our former notations (\ref{defin:Iab}) have been employed.  To
this end, it must be noted that $|4 \pi \tau \Delta \nu + 2\dot{\phi}
(0) + {\cal I} \Delta D| \geq 1$ is satisfied for all $\Delta \nu$ and
${\cal I}$, except those $ \Delta \nu $ which are in the vicinity of
the lineshape's maximum, $ \Delta \nu \simeq -
\dot{\phi}(0)/(4\pi\tau)$, and for $A_\omega$ such that $|{\cal I}
\Delta D| \lesssim 1$.  The latter condition happens to be more
restrictive as compared to the one assumed here. This is due the fact
that the diagonal matrix elements are usually much bigger than the
non-diagonal ones (see table \ref{tabD} for comparison).  By making
use of (\ref{C:simp}), (\ref{cond_omega}) and presuming that the above
condition is fulfilled, one can show that
\begin{eqnarray}
\fl  2 D_{b,a} {\cal I} \Im 
\left[ \int_{-\infty}^{x} \exp \left\{- \frac {\tau^2 u^2}{2 \sigma^2_{\rm t}} \right\}
\widetilde{C}(u,\Delta \nu,A_\omega,r) \rmd u  \right] \approx
\nonumber \\
\label{eq:exp}
\fl - \sqrt{2 \pi} D^2_{b,a} {\cal I}^3 \frac{\sigma_{\rm t}}{\tau} 
\frac{\Im(\Delta D)}
{\left[4 \pi \tau \Delta \nu + 2 \dot{\phi}(0) + {\cal I} \Re (\Delta D)\right]^2 +
{\cal I}^2 [\Im (\Delta D)]^2}
\left[ 1+{\rm erf} \left( \frac{\tau x }{\sqrt{2}\sigma_{\rm t}}\right) \right].
\end{eqnarray}
Note that, for any $x$, arbitrary laser parameters and $D_{b,a}$, this
expression is negative, as it should be, since $\Im(\Delta
D)=-\Im(D_{b,b}) =(4\pi \alpha)^{-1} \sigma_{b}^{(\gamma)}(\omega_{\rm
  L})>0$.  For small pulse energies considered here, the right--hand
side of (\ref{eq:exp}) is proportional to ${\cal I}^3$, which permits
the appropriate exponent in (\ref{eq:rho-Cappr}) to be expanded in
terms of $A_\omega$. On retaining two leading terms in this expansion,
this yields
\begin{eqnarray}
\fl \rho(A_\omega,r;[\phi]) = \frac{c}{4\pi}
        \frac{2}{\omega_{\rm L}} \sigma^{(\gamma)}_b (\omega_{\rm L}) {\cal I}
\int_{- \infty}^{+ \infty} \rmd \Delta \nu 
\int_{- \infty}^{+ \infty} \rmd x 
\exp \left\{- \frac {\tau^2 x^2}{2 \sigma^2_{\rm t}} \right\}
\left|\widetilde{C}(x,\Delta \nu,A_\omega,r)\right|^2
\nonumber \\
\label{exp_ser}
\fl \times \left\{
1 - \frac{\sigma_{\rm t}}{\tau} \frac{\sqrt{2 \pi} D^2_{b,a} {\cal I}^3  \Im(\Delta D)}
{\left[4 \pi \tau \Delta \nu + 2 \dot{\phi}(0) + {\cal I} \Re (\Delta D)\right]^2 +
{\cal I}^2 [\Im (\Delta D)]^2}
\left[ 1+{\rm erf} \left( \frac{\tau x }{\sqrt{2}\sigma_{\rm t}}\right) \right] \right\}.
\end{eqnarray}
Further calculations are straightforward but cumbersome.  Hence, some
technical details will be given below for the first term in the curly
brackets in (\ref{exp_ser}) only; the corresponding contribution is
denoted as $\rho^{(0)}(A_\omega,r;[\phi]) $. In this simplest case, by
substituting (\ref{C:simp}) for $\widetilde{C}(x)$ and interchanging
the order of integration, one arrives at
\begin{eqnarray}
\rho^{(0)}(A_\omega,r;[\phi]) =
\nonumber \\
\fl  \frac{c}{4\pi}
 \frac{\sigma^{(\gamma)}_b (\omega_{\rm L}){\cal I}^3}{\tau \omega_{\rm L}} D^2_{b,a} 
\int_{-\infty}^{+\infty} 
\exp\left\{ - \frac {\tau^2 x^2}{2 \sigma^2_{\rm t}} - 
\sqrt{2 \pi} \frac{\sigma_{\rm t}}{\tau} {\cal I} \Im(\Delta D) 
{\rm erf}\left( \frac{\tau x }{\sqrt{2}\sigma_{\rm t}}\right)  \right\} \rmd x
\nonumber \\
\lo{\times} \int_{-\infty}^{x} 
\exp\left\{ - \frac {\tau^2 u^2}{\sigma^2_{\rm t}} + 
\sqrt{2 \pi} \frac{\sigma_{\rm t}}{\tau} {\cal I} \Im(\Delta D) 
{\rm erf}\left( \frac{\tau u }{\sqrt{2}\sigma_{\rm t}}\right)  \right\} \rmd u
\nonumber \\
\label{eq:wood1}
\fl = \frac{c}{4\pi} 
\frac{\sigma_{\rm t} \sigma^{(\gamma)}_b {\cal I}^2}{\sqrt{2} \tau^2 \omega_{\rm L}} 
\frac{D^2_{b,a}}{\Im(\Delta D)} 
\left\{\sqrt{\frac{\pi}{2}} - \int_{-\infty}^{+\infty}
\exp\left[ - 2 u^2 - \sqrt{2 \pi} \frac{\sigma_{\rm t}}{\tau} {\cal I} \Im(\Delta D) 
{\rm erfc}(u)  \right]  \rmd u \right\} .
\end{eqnarray}
For two most important cases of interest, $|{\cal I} \Im(\Delta D) |
\ll 1$ and $|{\cal I} \Im(\Delta D) | \gtrsim 1$, the integral in
(\ref{eq:wood1}) can be evaluated, respectively, by expanding the
integrand into the power series in ${\cal I} \Im(\Delta D)$ and by
using the method of the steepest descent.  This yields
\begin{eqnarray}
\fl  \int_{-\infty}^{+\infty} 
\exp\left[ - 2 u^2 - \sqrt{2 \pi} \frac{\sigma_{\rm t}}{\tau} {\cal I}
\Im(\Delta D) 
{\rm erfc}(u)  \right]  \rmd u 
\nonumber \\
\label{step:desc}
\lo{\asymp} \left\{
\begin{array}[c]{ll} 
\sqrt{\pi/2} - \pi (\sigma_{\rm t} / \tau) \Im(\Delta D) {\cal I} 
       \qquad & {\rm for} \; |{\cal I} \Im(\Delta D) | \ll 1  \\ \ms
\left( \sqrt{\pi}  \tau^2 / 2^{3/2} \sigma^2_{\rm t} \right) 
                                \left[ \Im(\Delta D)  {\cal I} \right]^{-2} 
        \qquad & {\rm for}  \; |{\cal I} \Im(\Delta D) | \gtrsim 1
     \end{array}
   \right. .
\end{eqnarray}
Accordingly, we are left finally with 
\begin{eqnarray}
\rho^{(0)}(A_\omega,r;[\phi])  
\nonumber \\
\label{rho_0}
\fl = \frac{c}{4\pi} \sqrt{\frac{\pi}{2}}
\frac{\sigma_{\rm t} }{ \tau^2 \omega_{\rm L}} 
\sigma^{(\gamma)}_b (\omega_{\rm L}) D^2_{b,a} 
\times \left\{
        \begin{array}[c]{ll} 
        \sqrt{\pi} (\sigma_{\rm t}/\tau) {\cal I}^3 
        \qquad & {\rm for} \; |{\cal I} \Im(\Delta D) | \ll 1 \\ \ms
        [\sqrt{2} \Im(\Delta D)] ^{-1} {\cal I}^2
        \qquad & {\rm for}  \; |{\cal I} \Im(\Delta D) | \gg 1
        \end{array}
\right. .
\end{eqnarray}
Note that in either limiting case, the result is independent, in
accord with our initial statement, of particular form of the chirp
$\phi$.  To clarify this point, one should emphasize that the only two
simplifications made in the course of derivation were: (i) the use of
$\widetilde{C}(x)$ given by (\ref{C:simp}) and (ii) the replacement of
the appropriate exponent by a unity. Both approximations are fully
justifiable for $A_\omega:\:|A_\omega D_{b,a}(\omega_{\rm L})|
\lesssim 1$ assumed here. Obviously, the latter condition is
consistent with the {\em first} relation in (\ref{rho_0}), which has
been formerly obtained by means of a different technique (and cast in
a slightly different analytic form) in \cite{Woodman}. Apart from
being chirp--independent, this result demonstrates also that
$\rho^{(0)}(A_\omega,r;[\phi]) \propto A^3_{\omega}$, as opposed to
the next limiting case, that is, $|{\cal I} \Im(\Delta D) | \gtrsim
1$, where $\rho^{(0)}(A_\omega,r;[\phi]) \propto A^2_{\omega}$.  A
gradual decrease of the exponent is clearly evidenced by numerical
simulations carried out by us for muonium, while scanning sufficiently
broad $A_\omega$--domain. Some results of this numerical study are
discussed in more detail in section~\ref{sec:appl} (see figure
\ref{fig:power}, for example). These show, in particular, that
equations (\ref{rho_0}) describe $\rho^{(0)}(A_\omega,r;[\phi])$
qualitatively correctly for both low and moderately high pulse
energies.

In a similar manner one can also obtain appropriate contribution,
denoted $\rho^{(1)}(A_\omega,r;[\phi])$, coming from the second term
in the curly brackets in (\ref{exp_ser}). After tedious calculations
whose details will be given elsewhere, one ends up with the following
expression for a sum of two contributions:
\begin{eqnarray}
\rho^{(0)}(A_\omega,r;[\phi])  + \rho^{(1)}(A_\omega,r;[\phi]) 
\nonumber \\
\label{rho_01}
\fl = \left\{ 
        \begin{array}[c]{ll} 
\frac{c}{4\pi} \frac{\pi}{\sqrt{2}}
\frac{\sigma^2_{\rm t} }{ \tau^3 \omega_{\rm L}} 
\sigma^{(\gamma)}_b (\omega_{\rm L}) D^2_{b,a} {\cal I}^3  \left(
  1 - \sqrt{2} \pi \frac{\sigma^2_{\rm t} }{ \tau^2 } D^2_{b,a} {\cal I}^2 
  \right )       
  \qquad & {\rm for} \; |{\cal I} \Im(\Delta D) | \ll 1 \\ \ms
\frac{c}{4\pi} \frac{\sqrt{\pi}}{2}
\frac{\sigma_{\rm t} }{ \tau^2 \omega_{\rm L}} 
\sigma^{(\gamma)}_b (\omega_{\rm L}) 
\frac{  D^2_{b,a} }{\Im(\Delta D)} {\cal I}^2 \left(
  1 - \frac{4 \sqrt{2} \pi }{3}  \frac{\sigma^3_{\rm t} }{ \tau^3 }  D^2_{b,a} {\cal I}^2 \right)
        \qquad & {\rm for}  \; |{\cal I} \Im(\Delta D) | \gtrsim 1
        \end{array}
\right. .
\end{eqnarray}
Here, according to (\ref{cond_omega}), we have discarded the terms in
$\rho^{(1)}(A_\omega,r;[\phi]) $ containing the second and higher
order derivatives of $\phi(x)$. This simplification results in the
fact that entire equation (\ref{rho_01}) turns out to be completely
independent of the chirp.

The right--hand side of equation (\ref{cond_omega}) depends (through
${\cal I}$) on the radius $r$. To eliminate this dependence and to
facilitate, thereby, an adequate comparison with the results of
numerical calculations, the left-- and right--hand sides of
(\ref{cond_omega}) can be integrated over the entire $XOY$--plane
perpendicular to the direction of the laser beam's propagation
($z$--axis). A clear physical meaning of such an averaging procedure
is discussed in the comments following equation (\ref{W_avar}).

By virtue of the identity,
\[
\fl \int_{-\infty}^{+\infty}\!\!\int_{-\infty}^{+\infty}
{\cal I}^n \rmd x \rmd y =
\frac{2 \pi \sigma_{\rm r}^2}{n}
\left(\frac{2}{\pi c^2}\right)^{n/2}
\left(\frac{\tau A_\omega}{\sigma_{\rm t} \sigma^2_{\rm r}} \right)^n
\equiv \frac{2 \pi \sigma_{\rm r}^2}{n} \widetilde{\cal I}^n, \quad n=1, 2, \ldots,
\]
which defines the spatially--independent part of ${\cal I}$, denoted
as $\widetilde{{\cal I}}\equiv \sqrt{2/\pi c^2} \tau A_\omega/
\sigma_{\rm t} \sigma^2_{\rm r}$, this yields eventually the following
$r$--independent result:
\begin{eqnarray}
\rho(A_\omega;[\phi]) \approx 
\int_{-\infty}^{+\infty}\!\!\int_{-\infty}^{+\infty}
\left(\rho^{(0)}(A_\omega,r;[\phi])  + \rho^{(1)}(A_\omega,r;[\phi]) \right) \rmd x \rmd y
\nonumber \\
\label{rho_fin}
\fl = \left\{ 
        \begin{array}[c]{ll} 
\frac{c}{4\pi} \frac{\sqrt{2} \pi^2}{3}
\frac{\sigma^2_{\rm t} \sigma^2_{\rm r} }{ \tau^3 \omega_{\rm L}} 
\sigma^{(\gamma)}_b (\omega_{\rm L}) D^2_{b,a} \widetilde{{\cal I}}^3  \left(
  1 - \frac{ 3 \sqrt{2}\pi}{ 5} \frac{\sigma^2_{\rm t} }{ \tau^2 } D^2_{b,a}\widetilde{{\cal I}}^2 
  \right )       
  \qquad & {\rm for} \; |\widetilde{{\cal I}} \Im(\Delta D) | \ll 1 \\ \ms
\frac{c}{4\pi} \frac{\pi^{3/2}}{2}
\frac{\sigma_{\rm t} \sigma^2_{\rm r} }{ \tau^2 \omega_{\rm L}} 
\sigma^{(\gamma)}_b (\omega_{\rm L}) 
\frac{  D^2_{b,a} }{\Im(\Delta D)} \widetilde{{\cal I}}^2 \left(
  1 - \frac{2 \sqrt{2} \pi }{3}  \frac{\sigma^3_{\rm t} }{ \tau^3 }  
  D^2_{b,a} \widetilde{{\cal I}}^2 \right)
        \qquad & {\rm for}  \; |\widetilde{{\cal I}} \Im(\Delta D) | \gtrsim 1
        \end{array}
\right. .
\end{eqnarray}
\begin{figure}[t]
  \centering
  \includegraphics[height=7.5cm,width=7.5cm]{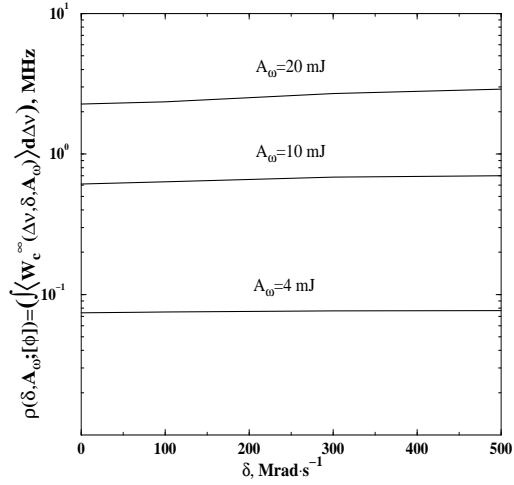}
\caption{The integral of the spatially--averaged two-step 3-photon 
  ionization probabilities of the ground state of muonium, $\langle
  W_c^{\infty}(\Delta \nu,\delta,A_\omega) \rangle $, over the entire
  range of the frequency detunings $\Delta \nu $, as a function of the
  chirp's magnitude, $\delta$; three particular pulse energies are
  considered: $A_\omega=4,\:10,\:20\:{\rm mJ}$.  The chirped frequency
  of the laser field employed is of the form: $\nu(t) \equiv \nu_{\rm
    L}+\nu_{\rm chirp}(t) \equiv \nu_{\rm L} + \frac{\delta}{4\pi }
  \left(1+\mbox{erf}(t/\tau)\right)$. The rest laser parameters used
  are: $\sigma_{\rm t}=51\:{\rm ns},\; \tau=120\:{\rm
    ns},\:\sigma_{\rm r}=0.64\:{\rm mm}$.}
\label{fig1c}
\end{figure} 

In order to check this equation, the numerical calculation of
$\rho(A_\omega;[\phi])$, in a wide range of $A_\omega$ and for typical
values of the rest physical parameters used in the 1S--2S experiment
in muonium, have been carried out. This was done by an extra numerical
integration of the 3--photon ionization line profiles $\langle
W_c^{\infty}(\Delta \nu,\delta,A_\omega) \rangle $ (see equation
(\ref{W_avar}) for definition) over the entire $\Delta \nu$--domain.
These results are shown in figure \ref{fig1c}, for three particular
pulse energies.  At $A_\omega=4\:{\rm mJ}$, being the only value
displayed which satisfies condition (\ref{cond_C}), the plot
demonstrates that $\rho(A_\omega;[\phi])$ is conserved to the 1\%
accuracy, within a wide range of the chirp's magnitude.  The mean
numerical value, $\rho(A_\omega=4\:{\rm mJ},\sigma_{\rm r}=0.64\:{\rm
  mm};[\phi]) \approx 0.080\:{\rm MHz}$, is to be compared with that
obtained by means of the {\em first} equation in (\ref{rho_fin}):
$\rho(A_\omega=4\:{\rm mJ},\sigma_{\rm r}=0.64\:{\rm mm};[\phi])
\approx 0.1242-0.0292=0.095\:{\rm MHz}$.  To make these results
comparable, the latter value includes also an additional weighting
factor, $(9\pi \sigma^2_{\rm r} )^{-1}$. Even though parameter
$\lambda$ is very close to unity: $\lambda =|\widetilde{{\cal I}}
\Im(D_{2s,2s}) |\simeq |\widetilde{{\cal I}} D_{2s,1s} | \approx 1$,
for $A_\omega=4\:{\rm mJ}$ and $\sigma_{\rm r}=0.64\:{\rm mm}$ used
here, numerical and analytical data are consistent to 20\% of relative
accuracy.  Interesting to note that the first/second relation in
(\ref{rho_fin}) overestimates/underestimate the true value of
$\rho(A_\omega;[\phi]) $. Much better agreement can be anticipated
(and actually takes place) for lower pulse energies and/or bigger
spatial dispersions $\sigma_{\rm r}$.  For $A_\omega=2\:{\rm mJ}$ and
$\sigma_{\rm r}=1.5\:{\rm mm}$ (apppropriate line profiles are
displayed in figure \ref{fig2}(a)), for example, numerical and
analytical results read: $\rho(A_\omega=2\:{\rm mJ},\sigma_{\rm
  r}=1.5\:{\rm mm};[\phi]) \approx 0.90\cdot 10^{-4}\:{\rm MHz}$ and
$(0.936-0.02)\cdot 10^{-4}=0.916\cdot 10^{-4}\:{\rm MHz}$,
respectively.

\subsubsection{\underline{High $A_\omega$}}
For higher pulse energies and/or smaller spatial dispersions
$\sigma_{\rm r}$, such that $|\widetilde{{\cal I}} \Im(\Delta D) | \gg
1$, equation (\ref{rho_fin}) fails to describe adequately the true
function $\rho(A_\omega;[\phi])$. This refers both to the absolute
values of $\rho(A_\omega;[\phi])$ and, particularly, to its dependence
on the magnitude of the chirp, which makes its appearance at high
pulse energies. It has been elucidated above that the main underlying
reason for this failure originates from the fact that ``the linear
approximation'' ceases to be valid any longer, due to corrections
caused by the $C^2$--term in equation (\ref{eq:full}).  This
non--linear term does not allow any reasonably accurate analytic
treatment to be developed, so that the numerical analysis of the
problem must be used instead.  To simulate the high--energy regime,
the results of our calculation are presented in figure \ref{fig1c},
for two pulse energies, $A_\omega=10,\:20\:{\rm mJ}$, and for typical
values of the rest physical parameters used in the 1S--2S experiment
in muonium. As in the case of low $A_\omega$, this was done by the
numerical integration of the 3--photon ionization line profiles
$\langle W_c^{\infty}(\Delta \nu,\delta,A_\omega) \rangle $ over the
entire $\Delta \nu$--domain.  The plots demonstrate in fact a
noticeable deviation of $\rho(A_\omega;[\phi])$ from a constant value,
with an increase of the pulse energy. For example, at
$A_\omega=20\:{\rm mJ}$ and for $\delta$ in the range $\delta=0 \ldots
500/(2\pi)\:{\rm MHz}$, this deviation (that is, the dependence on the
chirp's magnitude) amounts already to 30\%; in absolute units, the
data read: $\rho(A_\omega=20\:{\rm mJ},\sigma_{\rm r}=0.64\:{\rm
  mm};[\phi])=2.265\:{\rm MHz}$ for $\delta=0\:{\rm MHz}$ and
$\rho(A_\omega=20\:{\rm mJ},\sigma_{\rm r}=0.64\:{\rm
  mm};[\phi])=2.90\:{\rm MHz}$ for $\delta=500/(2\pi)\:{\rm MHz}$.
Furthermore, the gauge of the given deviation may be expected to be
even more pronounced at higher pulse energies, thus preventing us
actually from viewing $\rho(A_\omega;[\phi])$ as a conserving quantity
in this case.  Although such $A_\omega$ are of no practical interest
at the moment for the $1S$--$2S$ experimental study with muonium, the
entire effect might be of relevance for future experimental studies
where much higher laser intensities are involved.

\section{Application to the experiment in muonium}
\label{sec:appl}

Let us apply the results obtained in preceding sections to particular
laser parameters adopted in the 1S-2S experiment in muonium
\cite{Maas}, by assuming that $|a\rangle \equiv 1S,\:|b\rangle \equiv
2S$ and setting $\phi(t) \equiv \delta t \left(1+ {\rm
    erf}(t/\tau)\right)/2$. It is convenient in this section to
measure the energies in the units: $\hbar=e^2=m^\ast=1$, where
$m^\ast=m_\e/(1+m_\e/m_\mu)\approx (1.00484)^{-1}\:m_\e$ stands for
the reduced mass of the electron and muon. Here, the value for the
electron to $\mu^-$ mass ratio, $m_\mu/m_\e=206.768262$, was used.
The quantity $a^\ast\equiv\hbar^2/(m^\ast e^2)=(m_\e/m^\ast)a_0$
stands for the unity of distance, with $a_0=\hbar^2/(m_\e \e^2)=0.529
\cdot 10^{-8}\:\mbox{cm}$ being the Bohr radius. Accordingly,
one--particle energies and dynamic polarizabilities,
$\alpha_{b,a}^{ij}(\omega)$, are measured in the units of $e^2/a^\ast$
and $(a^\ast)^3$.

As was mentioned above, 3--photon ionization probability,
$W_c^{\infty}(\Delta \nu,\delta,A_\omega,r)$, can be obtained by
accurate numerically solving either the system
(\ref{ala2})-(\ref{cont}) or (\ref{r2s})-(\ref{r_cont}).  In both
cases this has been done by the stepwise time integration, starting
from $t_{0}=-3\sigma_{\rm t}$, while scanning sufficiently broad
$\Delta \nu$--domain centered at $\Delta \nu=0$ (i.e. $\omega_{\rm
  L}=\frac{1}{2} \omega _{2s,1s}$), for all $r \equiv \sqrt{x^2+y^2}$
on the $r$--mesh: $r_k=k \sigma_{\rm r}/10,\; k=0 \ldots 50$.
Numerical values of the matrix elements, $D_{b,a},\;(a,b)=1s,2s$,
which enter (\ref{ala2})-(\ref{cont}) were calculated in
\cite{Bass,Beausoleil,YJ}; these are compiled in table \ref{tabD},
together with the value of the single photoionization cross section of
the $2S$--level at $\omega=\omega_{\rm L}$. The latter is given
explicitly by (see \cite{YJ} and references therein)
\begin{equation}
\label{sigma2s}
\fl \sigma_{2s}^{(\gamma)}
(\omega_{\rm L}) = \frac{2^{14}\pi^2}{3}\alpha 
\left(1+3\frac{I_{2s}}{\omega_{\rm L}}\right)
\left(\frac{I_{2s}}{\omega_{\rm L}}\right)^4
\frac{\e^{-4 \eta \arctan(2/\eta)}}{1-\e^{-2\pi\eta}},\qquad
\eta = \sqrt{\frac{4I_{2s}}{\omega_{\rm L}- I_{2s}}}
\end{equation}
where $\alpha$ is the fine structure constant and $I_{2s}=1/8$ denotes
the ionization potential of the $2S$--state.

\begin{table}[t]
\caption{The values of the two--photon matrix elements
$D_{b,a},\;(\mbox{b,a})=1s,2s$, defined by the sums in equations
(\ref{Vba})-(\ref{Vbb}), and the value of the 1--photon photoionization
cross section of the $2S$--level, given by (\ref{sigma2s}), at
$\omega_{\rm L} \approx 3/16\:a.u.\;(\lambda_{\rm L} =
244\:\mbox{nm})$; linear polarization of the photon, $\bepsilon$, is
assumed in both cases. }
\begin{indented}
\item[]
\[
\begin{array}{cccc} \br 
D_{1s,1s}(\omega_{\rm L}) & D_{2s,2s} (\omega_{\rm L}) &
D_{2s,1s}(\omega_{\rm L}) & \sigma_{2s}^{(\gamma)} (\omega_{\rm L}) \\
\mr \ms -5.7141 & 29.8535 -\rmi \: 12.8232 & 7.8535 & 0.2205 \\ \ms \br
\end{array}
\]
\label{tabD}
\end{indented}
\end{table}

At the final stage of our numerical procedure, a spatial averaging has
been performed in order to obtain $r$-independent ionization profiles,
$\langle W_c^{\infty}(\Delta \nu,\delta,A_\omega) \rangle$.  This has
been achieved by an extra integration of $W_c^{\infty}(\Delta
\nu,\delta,A_\omega,r)$ over entire $XOY$--plane perpendicular to the
direction of the laser beam's propagation:
\begin{equation}
\label{W_avar}
\langle W_c^{\infty}(\Delta \nu,\delta,A_\omega) \rangle  
\equiv \frac{1} {9\pi \sigma_{\rm r}^2}
\int_{-\infty}^{+\infty} \!\!\!\int_{-\infty}^{+\infty}
W_c^{\infty}(\Delta \nu,\delta,A_\omega,r)\: \rmd x \rmd y.
\end{equation}
Basically, this newly introduced quantity can be interpreted as an
averaged ionization probability related to the entire laser beam spot,
since $\langle W_c^{\infty}(\Delta \nu,\delta,A_\omega) \rangle $ is
independent of the distance from the beam's axis. Alternatively, in
view of the low values of the muon density in the media \cite{Maas}
(see below), $\langle W_c^{\infty}(\Delta \nu,\delta,A_\omega) \rangle
$ can be called as ``the ionization probability per one (muon) atom in
the beam''.

An auxiliary factor, $1/(9\pi \sigma^2_{\rm r})$, has been introduced
in (\ref{W_avar}) in order to make $\langle W_c^{\infty}(\Delta
\nu,\delta,A_\omega) \rangle$ dimensionless, as the original
probability $W_c^{\infty}(\Delta \nu,\delta,A_\omega,r)$ is.  This
particular choice was adopted according to the so--called
\mbox{``$3\sigma$''}--rule being inherent to various problems
involving Gaussian law. In the case considered, the
``$3\sigma$''--rule makes its appearance through the fact that the
values of $W_c^{\infty}(\Delta \nu,\delta,A_\omega,r)$ happen to be
almost negligible for $r>3\sigma_{\rm r}$, unless the pulse energies
higher than $A_\omega =20\:{\rm mJ}$, say, are considered.  For such a
high energy regime, the probability $W_c^{\infty}(\Delta
\nu,\delta,A_\omega,r)$ has a long--distance ``tail'' spreading out
beyond the {\em effective} beam radius, $r = 3\sigma_{\rm r}$, within
which the values of $W_c^{\infty}(\Delta \nu,\delta,A_\omega,r)$ may
be very close to, yet less than, $1$.  In this situation the proper
normalization factor different from that chosen above must be used so
as to prevent $\langle W_c^{\infty}(\Delta \nu,\delta,A_\omega)
\rangle $ from being bigger than $1$.  As is clearly evidenced by our
calculations, the values of $\langle W_c^{\infty}(\Delta
\nu,\delta,A_\omega) \rangle$ defined by equation (\ref{W_avar})
exceed 1 starting already from $A_\omega \simeq 20 \div 25\:{\rm mJ}$,
depending on the rest laser parameters.  One should point out,
however, that such $A_\omega$ are hardly attainable currently with
proper pulsed lasers sources in the required frequency range.  This
makes the high energy regime in the resonant 3--photon ionization with
muonium to be mostly of academic rather than practical interest at the
moment, thus justifying the definition of equation (\ref{W_avar}).  To
avoid any misunderstanding, it must be clearly stated once again that
the spurious effect mentioned here originates solely from the
particular choice of the normalization factor in (\ref{W_avar}), and
has, therefore, nothing to do with either the model developed in the
work or a lack of the numerical accuracy employed by us. The latter
was chosen to be equal to $10^{-7}$ which proves to be sufficient to
ensure that $W_c^{\infty}(\Delta \nu,\delta,A_\omega,r) \leq 1$ holds
(as it should!)  for any particular individual set of all parameters
involved in the problem, including $r,\;\Delta \nu,\;\delta,\;\tau$,
and $\sigma_{\rm r}$.

\begin{figure}[hb]
  \centering \includegraphics[height=7.5cm,width=14cm]{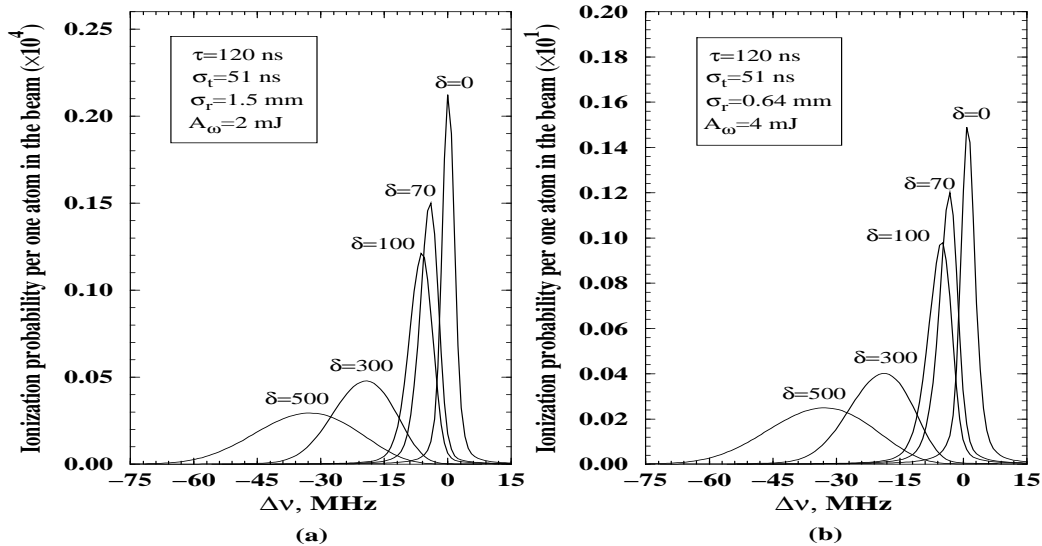}
\caption{Two-step 3-photon ionization probabilities, 
  $\langle W_c^{\infty}(\Delta \nu,\delta,A_\omega) \rangle $, of the
  ground state of muonium by the laser field with the chirped
  frequency $\nu(t) \equiv \nu_{\rm L}+\nu_{\rm chirp}(t) \equiv
  \nu_{\rm L} + \frac{\delta}{4\pi }\left(1+{\rm
      erf}(t/\tau)\right),\;\delta =500,300,100,70,0\:{\rm Mrad}\cdot
  {\rm s}^{-1}$, versus laser frequency detuning, $\Delta \nu \equiv
  \left(\omega_{\rm L}-\frac{1}{2}\omega_{2s,1s}\right)/2\pi $. Two
  pairs of different values for the pulse energy, $A_\omega$, and the
  spatial dispersion of the laser signal, $\sigma_{\rm r}$, are used:
  (a) $A=2\:{\rm mJ},\; \sigma_{\rm r} = 1.5 \:{\rm mm}$, and (b)
  $A=4\:{\rm mJ}, \sigma_{\rm r} = 0.64 \:{\rm mm}$.}
\label{fig2}
\end{figure}

The averaged probability $\langle W_c^{\infty}(\Delta
\nu,\delta,A_\omega) \rangle$ readily enables one to estimate the
expected number of experimentally observed ionization events, $N_{\rm
  event}$, defined as the number of the muonium atoms which are
ionized by a sequence of $N_{\rm shot}$ identical Gaussian laser
pulses (\ref{intens}) within the time $T$ and detected eventually in
apparatus.  Such a formulation of the problem corresponds direct to
the actual experimental situation where the gas media containing
muonium atoms interacts with $N_{\rm shot}=25$ laser pulses per
second.  Under these conditions, the required number of the ionization
events to be detected during the observation time $T \gg N_{\rm shot}
\tau$ can be estimated by means of the following simple relation:
\[
N_{\rm event}=\rho_0 \cdot S \cdot L \cdot T \cdot \eta \cdot N_{\rm
  shot} \cdot W_c^{\rm max}(\delta,A_\omega).
\]
Here, $\rho_0 \simeq 4 \cdot 10^{-3}\:{\rm atoms/mm}^3$ is the spatial
density of the muonium atoms in the chamber, $S=9\pi \sigma^2_{\rm
  r}\simeq 11\:{\rm mm}^2$ and $L \simeq 10^3\:{\rm mm}$ stand,
respectively, for the {\em effective} cross section of the laser beam
(see above) and the total path which this beam travels in the media,
$\eta \simeq 10\%$ denotes an efficiency of registration of the
muonium atoms in apparatus; finally, $W_c^{\rm max}(\delta,A_\omega)$
is the value of the averaged 3--photon ionization probability $\langle
W_c^{\infty}(\Delta \nu,\delta,A_\omega) \rangle$ at its maximum, for
particular values of the chirp and the energy of the pulse.  Provided
$\delta$ and $A_\omega$ are fixed, this maximum is achieved
approximately at $\Delta \nu \approx -\delta/(4 \pi)$, almost
independently of $A_\omega$, as can be seen in figure \ref{fig2}(b).
In particular, $W^{\rm max}_c(\delta=10\:{\rm MHz},A_\omega = 4\:{\rm
  mJ}) \approx 0.014$ thus leading, for example, to the following
expected number of events detectable during the observation time
$T=3600\:{\rm s}$: $N_{\rm event}\simeq 5800$.  It is rather
instructive that this value happens to be quite close to that found
preliminary in the $1S-2S$ experiment in muonium being currently
underway.
\begin{figure}[t]
  \centering
  \includegraphics[height=7.5cm,width=14cm]{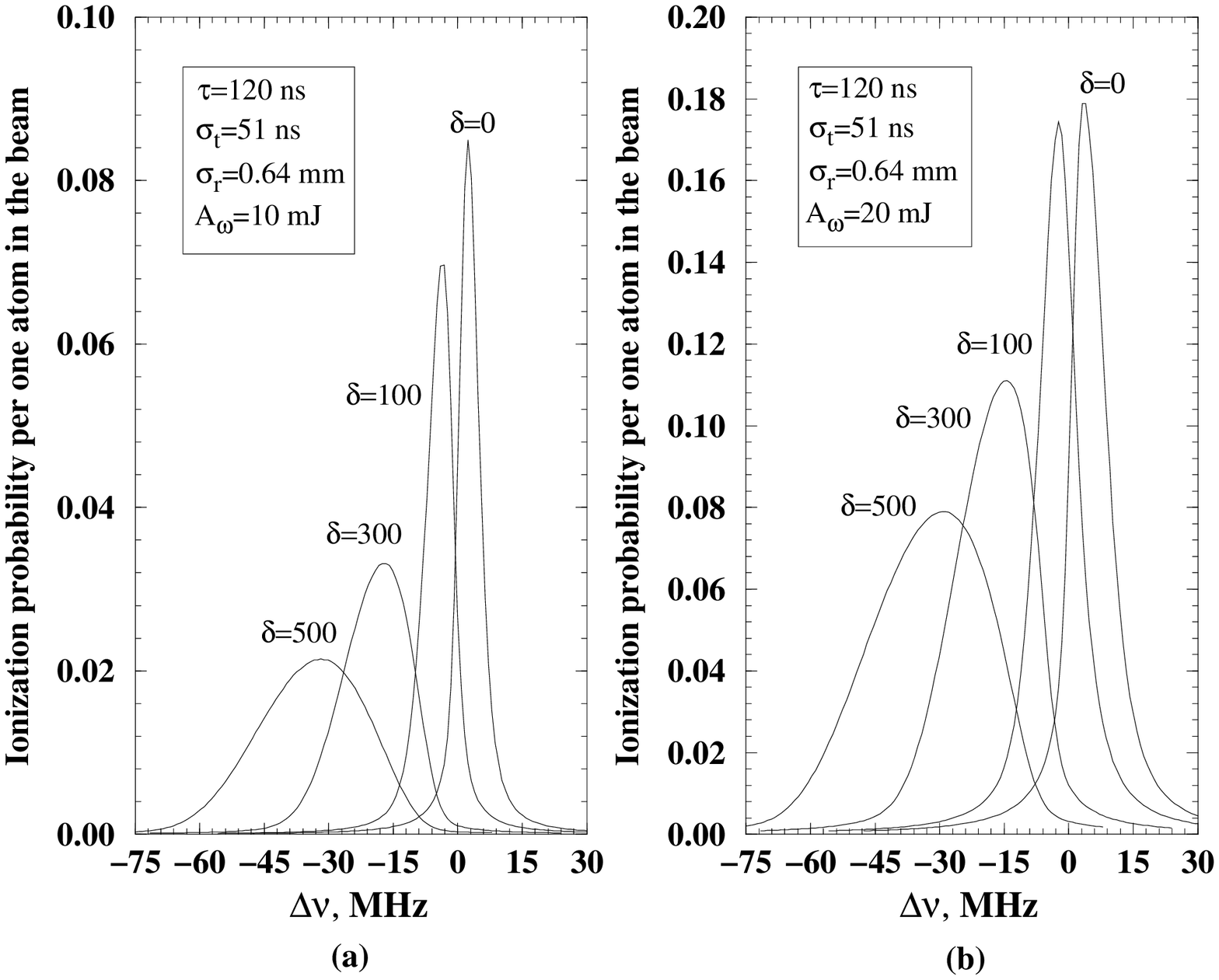}
\caption{Two-step 3-photon ionization probabilities, 
  $\langle W_c^{\infty}(\Delta \nu,\delta,A_\omega) \rangle $, of the
  ground state of muonium by the laser field with the chirped
  frequency $\nu(t) \equiv \nu_{\rm L}+\nu_{\rm chirp}(t) \equiv
  \nu_{\rm L} + \frac{\delta}{4\pi }
  \left(1+\mbox{erf}(t/\tau)\right),\; \delta =0,100,300,500\:{\rm
    Mrad}\cdot {\rm s}^{-1}$, versus laser frequency detuning, $\Delta
  \nu \equiv \left(\omega_{\rm L}-\frac{1}{2}
    \omega_{2s,1s}\right)/2\pi $, for two pulse energies: (a)
  $A_\omega=10\:\mbox{mJ}$ and (b) $A_\omega=20\:\mbox{mJ}$. }
\label{fig3}
\end{figure} 
\begin{figure}[b]
  \centering \includegraphics[height=7cm,width=14cm]{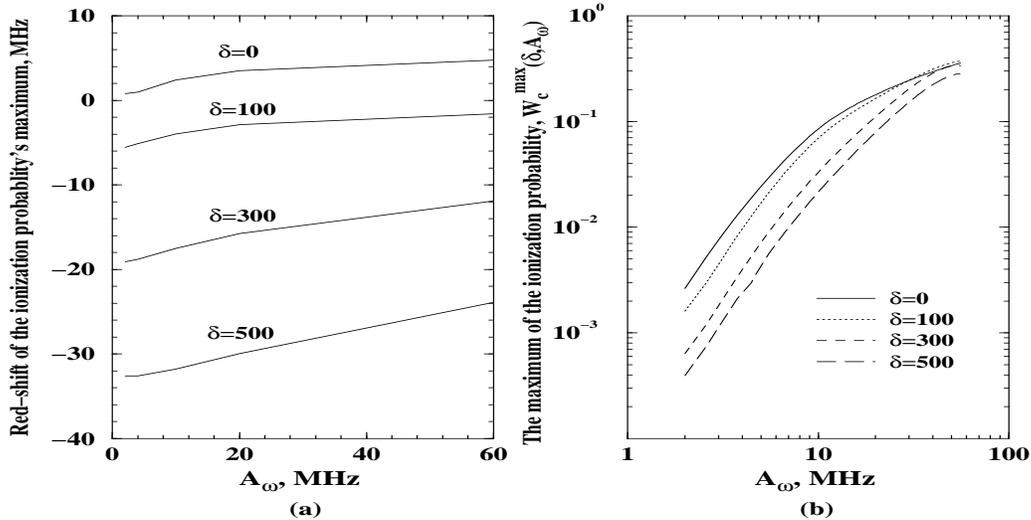}
\caption{(a) - the frequency red--shift of the photoionization
  profile's maximum, against the pulse power $A_\omega$. The chirped
  frequency of the laser field is assumed to be of the form: $\nu(t)
  \equiv \nu_{\rm L}+\nu_{\rm chirp}(t) \equiv \nu_{\rm L} +
  \frac{\delta}{4\pi } \left(1+\mbox{erf}(t/\tau)\right),\; \delta
  =0,100,300,500\:{\rm Mrad}\cdot {\rm s}^{-1}$; (b) - the
  power--dependence of the line profile's maximum value. In both cases
  the rest laser parameters are the same as in figures
  \ref{fig3}(a,b). }
\label{fig4}
\end{figure} 

\begin{figure}[t]
  \centering \includegraphics[height=7cm,width=7cm]{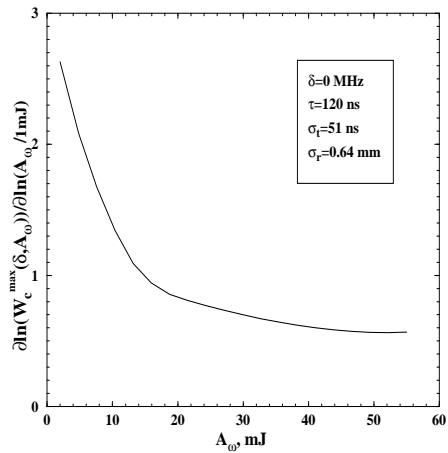}
\caption{The $A_\omega$--dependence of the exponent, $\mu(A_\omega)$,
  defined by the relation: \\$W_{c}^{\rm max} (\delta,A_\omega) =
  \gamma(\delta) A^{\mu(A_\omega)}$.}
\label{fig:power}
\end{figure}

\begin{figure}[t]
  \centering \includegraphics[height=7cm,width=15cm]{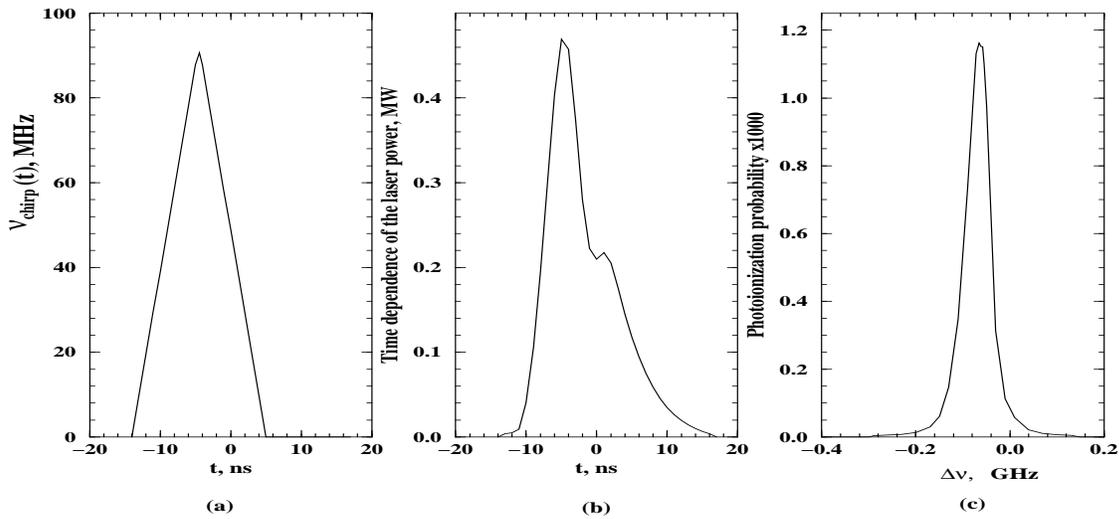}
\caption{Time--dependence  of the chirp -- (a) -- and the laser 
  power -- (b) -- both detected in the former experimental
  measurements of the $1S-2S$ separation in muonium
  \protect{\cite{Maas}}; (c) - photoionization line profile measured
  in \protect{\cite{Maas}}, versus laser frequency detuning, $\Delta
  \nu \equiv \left(\omega_{\rm L}-\frac{1}{2}
    \omega_{2s,1s}\right)/2\pi $. }
\label{fig5}
\end{figure}

Some results of our simulations are presented in figures
\ref{fig2}-\ref{fig5}, for various magnitudes of the chirp $\delta$
and various parameters of the laser pulse which are typically used in
the measurements of the $1S-2S$ energy separation in muonium;
particular values employed are indicated in figures
\ref{fig2},\ref{fig3}. Both these graphs demonstrate appreciable
red--shift of the maxima (i.e.  towards lower frequencies, relative to
$\nu_{\rm L} = \omega_{2s,1s}/4\pi$) of the photoionization profiles
with an increase of the chirp.  It must be noted, however, that this
shift is caused by a combination of two competing mechanisms: (i) the
shift being due to the combined {\em AC Stark} effect for the $1S$-
and $2S$--levels and (ii) that arising because of the chirp itself.
The curves show that the former mechanism turns out to be small
compared with the latter at relatively small pulse energies,
$A_\omega=2 \ldots 4\:\mbox{mJ}$.  At $A_\omega=4\:\mbox{mJ}$ and
$\delta=0$, for instance, i.e. in the case of completely unchirped
laser signal, the {\it AC Stark\/} shift amounts approximately to
$1\:\mbox{MHz}$.  One can anticipate, however, that the relative
contribution of the {\it AC Stark\/} shift will increase as the laser
intensity increases, since its magnitude is linearly proportional to
the laser intensity, whereas the chirp--induced shift is almost
intensity--independent. This behaviour can be seen in figures
\ref{fig3}(a,b) and is summarized in figure \ref{fig4}(a) showing, in
particular, that the {\it AC Stark\/} shift becomes equals to
$4\:\mbox{MHz}$ at $A=20\:\mbox{mJ},\;\delta=0$. In addition to the
shift of the maxima, the curves in figures \ref{fig2},\ref{fig3}
exhibit a strong dependence on both the spatial dispersion
$\sigma_{\rm r}$ and the power $A_\omega$ of the laser signal.  It
should be noted that an appreciable distortion of the photoionization
line shapes happens to be particularly enhanced at relatively large
chirp values, where the line profiles become asymmetric. Generally
speaking, this asymmetry is an intrinsic feature of the 3--photon
resonant photoionization occurring under the action of the {\em
  chirped} pulsed laser signal.  Moreover, an asymmetry of the
ionization line shapes happens to be one of the most pronounced
manifestations of the non-monochromaticity of the laser field. This
phenomenon is basically due to the fact that the chirp, even though
its relative magnitude amounts to only $\delta/\omega_{\rm L} \simeq
10^{-8}\ldots 10^{-7}$ in our calculations, violates an equivalence of
initial ($t=-\infty$) and final ($t=+\infty$) time moments. Indeed,
the photoionization probability turns out to be either more efficient
or suppressed, depending on whether the time--dependent chirped laser
frequency is in the resonance with the 2--photon $|a\rangle
\rightarrow |b\rangle$ transition or slightly off it. The process as a
whole becomes then somewhat ``time--sensitive'', although the form of
the laser pulse envelope used is time--invariant.  It should be noted,
however, that the gauge of the above asymmetry depends strongly on the
laser power, as well as particular forms of the laser amplitude and
the chirp.  This dependence can be seen in figures \ref{fig3}(a,b)
where higher $A_\omega$--values as compared with figures
\ref{fig2}(a,b) are used.  The former show that the asymmetry
discussed is less pronounced as long as higher laser intensities are
involved. Two essential characteristics of the line profiles' maximum,
that are, the shift and the magnitude, are plotted in figure
\ref{fig4}(a,b) as functions of the laser power for various values of
the chirp.  In particular, the graph \ref{fig4}(b) provides
information about the dependence of the line shapes' maximum value,
$W_{c}^{\rm max}(\delta,A_\omega)$, on the pulse energy $A_\omega$.
It is quite natural to parameterize this relation by the following
simple power law: $W_{c}^{\rm max}(\delta,A_\omega) = \gamma(\delta)
A^{\mu(A_\omega)}$.  The exponent here proves to be a slow--varying
function of the pulse energy, such that $(\rmd \mu(A_\omega)/\rmd
A_\omega)/ \mu(A_\omega) \ll (A_\omega \ln A_\omega)^{-1}$.  According
to this condition, $\mu(A_\omega)$ can be estimated as
\[
\mu(A_\omega) \approx \frac{\partial}{\partial (\ln A_\omega)} \ln
\left[W_{c}^{\rm max} (\delta,A_\omega)\right].
\]
One should note, for example, an appreciable deviation of the
$\mu(A_\omega)$--values from $3$ which must be expected for a 2--step
3--photon ionization of an atom in the state $|a\rangle$ by a {\em
weak, monochromatic and spatially homogeneous laser field}, without
the relaxation of the intermediate levels.  For $\delta=0$, the
$A_\omega$--dependence of $\mu(A_\omega)$ is displayed in
figure~\ref{fig:power}.  This graph demonstrates that
$\mu(A_\omega=2\:{\rm mJ}) \approx 2.75$ gradually decreases as the
pulse energy increases and tends, for $A_\omega \gg 20\:{\rm mJ}$, to
an almost constant value, $\mu \approx 0.5$, thus indicating on the
presence of saturation in the 3--photon transition considered.

Finally, it should be mentioned that the above consideration has been
developed mainly in the attempt to simulate those line profiles of the
two--step 3--photon ionization probabilities that are supposed to be
measured soon in the new $1S-2S$ experimental investigation of
muonium. Essential physical parameters employed in this study are
still not fixed completely and are subject to further changes. It is
therefore rather tempting to apply our present technique to
appropriate data formerly collected within the framework of the
previously employed experimental setup~\cite{Maas}. These are shown in
figures\ \ref{fig5}(a,b), along with the photoionization line profile
obtained within the framework of the current theoretical approach,
figure\ \ref{fig5}(c).  The fact that the corresponding experimental
line shape (not shown) is virtually indistinguishable from the
theoretical one provides an additional and independent check of the
overall validity of the model developed in this work.

\section{Conclusion}

We have developed a simple theoretical scheme intended to describe, to
the $1\:\mbox{MHz}$ accuracy, the stepwise 3--photon resonant
photoionization in hydrogenic systems, induced by the chirped laser
field with time--dependent amplitude. It has been shown that such an
accuracy can be achieved within the framework of a simple 3-level
model, by taking into account (i) the {\it AC Stark} shifts and (ii)
non-zero ionization rates of the levels involved, together with (iii)
a spatial inhomogeneity of the laser signal and (iv) arbitrary
$t$-dependencies of its intensity (pulse envelope), $I(\bi{r},t) $,
and frequency, $\omega(t)$.  The system of equations
(\ref{ala2})-(\ref{cont}) or, equivalently, equations
(\ref{r2s})-(\ref{r_cont}) is a key point of the method employed.
These are of independent significance as being not specifically
restricted to particular states of reference, laser intensities, and
chirps, so that the results obtained for $1S-,\:2S-$ and $\varepsilon
P$--states can be generalized on arbitrary
$|a\rangle,\:|b\rangle,\:|c\rangle$ levels of hydrogen--like atoms.
Excited $ns-$levels, such that $2\leq n \leq 5$, are of particular
current interest for ultra-high precision laser spectroscopy, and it
should be expected that an adequate interpretation of experimental
data should require an accurate theoretical account of a wide spectrum
of light--induced effects. Some of these effects are beyond the scope
of our present consideration and, in the first instance, comprise
appropriate corrections due to inevitable motion of atoms in the media
and associated {\it second\/} order Doppler shifts, as being of major
importance. Indeed, even for CW lasers where the signals are usually
almost unchirped in the laboratory frame, the amplitude and the
frequency of the pulsed laser signals become essentially
time--dependent in the atomic rest frame, thus leading, as was
demonstrated above, to the shifts and distortions of
ionization/excitation lines. The relative contribution of the effects
discussed is estimated as $\alpha^2 v^2_z$ with $v_z$ being the atomic
velocity in the laboratory frame, and these should be incorporated in
the theoretical scheme developed here, along with
relativistic/radiative corrections $\simeq \alpha^2$ to the operator
of the particle--laser field interaction. This work is currently in
progress, and we consider it as a subject of forthcoming publications.

       
\ack The authors are indebted to G. zu Putlitz for his constant
support and encouragement. This work has been strongly motivated and
influenced by discussions within the muonium 1S-2S collaboration,
particularly with P.E.G.  Baird, M.G. Boshier, P.G.H. Sandars, and
W.T. Toner.  One of us (V.Y.), wishes to acknowledge his gratitude to
Prof. I.P. Grant (FRS) for his kind advice and to the
Volkswagen--Stiftung and the Royal Society for financial support.
Also, this work has been funded in part by the grants from NATO, the
Royal Society and the Bundesminister f\"{u}r Bildung und Forschung of
Germany.

\end{document}